\documentclass[aps,prx,twocolumn,showpacs,amsmath,amssymb,superscriptaddress]{revtex4-2}
\usepackage{amsmath, amsthm, amsfonts, amssymb,amstext}
\usepackage[utf8]{inputenc}
\usepackage[english]{babel}
\usepackage[T1]{fontenc}
\usepackage{hyperref}
\usepackage{standalone}
\usepackage{import}
\usepackage{multirow}
\usepackage{diagbox}
\usepackage{adjustbox}
\usepackage{braket}
\usepackage[caption=false]{subfig}

\usepackage[external]{forest}
\usepackage{forest}
\usepackage[table]{xcolor}
\definecolor{myblue}{rgb}{0.2,0.2,0.8}
\definecolor{myred}{rgb}{1,0.,0.3}
\usepackage{hyperref}
\hypersetup{
  colorlinks   = true, 
  urlcolor     = blue, 
  linkcolor    = blue, 
  citecolor   = blue
}

\def\beq{\begin{equation}}
\def\eeq{\end{equation}}
\def\barray{\begin{eqnarray}}
\def\earray{\end{eqnarray}}

\begin{document}
\title{Optimizing bias-tailored quantum  error correction beyond code-capacity noise}
\author{C. Benito}
\affiliation{Instituto de F\'isica Te\'orica UAM-CSIC, Universidad Aut\'onoma de Madrid, Cantoblanco, 28049, Madrid, Spain}
\author{I. J. Vel\'azquez-Res\'endiz}
\affiliation{Instituto de F\'isica Te\'orica UAM-CSIC, Universidad Aut\'onoma de Madrid, Cantoblanco, 28049, Madrid, Spain}
\author{A. Bermudez}
\affiliation{Instituto de F\'isica Te\'orica UAM-CSIC, Universidad Aut\'onoma de Madrid, Cantoblanco, 28049, Madrid, Spain}
\begin{abstract}
We find that the substantial advantages predicted for bias-tailored quantum error correction (QEC) under code-capacity noise are strongly reduced once realistic syndrome extraction and circuit-level noise models are considered. We start by comparing XZZX codes to rectangular surface codes with a bias-dependent optimised anisotropy. Although code-capacity simulations predict an advantage of rectangular surface codes in the limit of high noise bias, this actually disappears under circuit-level noise, making the XZZX codes the preferred and simplest  choice even for platforms that allow for a flexible  variation of the code layout  adapted to changes in noise calibration. Our results identify bias degradation during syndrome extraction under circuit-level noise as the central limitation of biased-tailored QEC. To partially mitigate this effect, we introduce a bias-filtering CNOT gadget that temporarily encodes the ancillary target qubit during syndrome extraction in a repetition code and, upon measurement and feed forward,  manages to reduce the bias degradation. In a regime of high-bias and low-idle errors, this bias-filtering gadget yields a few-percent relative improvement of   the XZZX code error threshold, demonstrating that lightweight  bias-filtering strategies can recover part of the lost bias-tailoring advantage for realistic circuit-level noise.
\end{abstract}
\maketitle

\setcounter{tocdepth}{0}
\begingroup
\hypersetup{linkcolor=black}
\tableofcontents
\endgroup

\section{\bf Introduction}
Quantum error correction (QEC)~\cite{PhysRevA.52.R2493,10.1098/rspa.1996.0136} is widely regarded as an essential ingredient for scalable quantum computation. It enables the suppression   of physical errors  through redundant  encoding and  conditional correction operations based on frequent syndrome readout and decoding~\cite{Shor1996,PhysRevA.57.127},  preventing the   amplification of errors through consecutive gates in the circuits through the use of fault-tolerant (FT)  design principles~\cite{aharonov1999faulttolerantquantumcomputationconstant}. As  the technology progresses reducing physical errors even further and allowing for ever larger levels of redundancy,   noisy quantum computers are expected to traverse the path from current intermediate-scale  prototypes to truly FT devices capable of operating  at the  megaquop regime~\cite{Eisert:2025ytq}  with thousands of logical qubits and $\mathcal{O}(10^6)$ gates, and ultimately  entering the teraquop  regime in which the circuit complexity with $\mathcal{O}(10^{12})$ gates can unlock   the full advantage of various quantum algorithms~\cite{Dalzell_McArdle}.
In recent years, we have witnessed a substantial progress through the demonstration of key  QEC building blocks, including the  exponential suppression of logical errors by continuously increasing  redundancy when physical operations lie  below the QEC threshold~\cite{Google2025below,Lacroix2025}, or  the demonstrations of a logical universal gate set and first small quantum  algorithms at the logical level~\cite{Postler_2022,SalesRodriguez2025,doi:10.1126/science.adp6016,mayer2024benchmarkinglogicalthreequbitquantum,Bluvstein2026,perlin2026faulttolerantexecutionerrorcorrectedquantum}.

\begin{figure}
\centering
\subfloat[]{\includegraphics[scale=4]{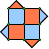}}
\subfloat[]{\includegraphics[scale=4]{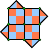}}\\
\subfloat[]{\includegraphics[scale=4]{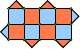}}
\caption{{\bf Qubit and stabilizer arrangement in the surface code}: data qubits are placed in the vertices of a rectangular lattice. Each face is associated to a stabilizer, which has $X$ (red) or $Z$ (blue) support on its vertices. a) $d=3$ CSS surface code. b) $d=3$ XZZX code. c) Anisotropic $(d_X,d_Z)=(3,5)$ CSS surface code}
\label{fig:surface-codes}
\end{figure}

 These  experiments have mainly focused on  topological codes~\cite{KITAEV20032,RevModPhys.87.307}, such as  surface~\cite{bravyi1998quantumcodeslatticeboundary,10.1063/1.1499754} and  color~\cite{PhysRevLett.97.180501,landahl2011faulttolerantquantumcomputingcolor} codes, in order  to benefit from their local stabilizer structure and relatively simple syndrome-extraction circuits, which become particularly advantageous in platforms with limited connectivity of the  entangling gates~\cite{PhysRevA.86.032324,Benito2025comparativestudyof} (see Fig.~\ref{fig:surface-codes}a for the surface code). The predicted performance of these QEC codes, quantified in terms of  their QEC error thresholds~\cite{aharonov1999faulttolerantquantumcomputationconstant,10.1098/rspa.1998.0166} or their resource overheads quantified in the  physical-to-logical qubit footprint required to enter the mega(tera)-quop regimes,  can strongly vary depending on how accurately the noise in the device is modeled. Typical noise models range  from code-capacity models that assume a perfect syndrome readout, to phenomenological noise models where this readout can fail but the corresponding faults do not propagate through the circuits and, ultimately, to full circuit-level noise models that incorporate noisy gates, correlated faults, and circuit-aware error propagation. These different noise models lead to widely different estimates of the above logical performance metrics for the same QEC code~\cite{Wang:2009bfa,landahl2011faulttolerantquantumcomputingcolor}.
 
 Moreover, important differences can even arise  within the same type of noise model,  as the way in which the errors affect the qubits, the syndrome readout, or the quantum gates  can incorporate a variety of microscopic details that typically lead to more structured effects within the same type of noise model.
A particularly  form of structured noise that is quite common in the literature of QEC is that of partially-biased dephasing, where phase-flip $Z$ errors occur at significantly higher rates than bit-flip $X$ ones. Such noise asymmetries naturally arise in several experimental  platforms, including trapped ions~\cite{PhysRevLett.123.110503,seis2023improving,ly5s-tjk2}, silicon spin qubits~\cite{takeda2022quantum,hetenyi2024tailoring,noiri2022fast,steinacker2025industry}, and certain cat-qubit architectures~\cite{lescanne2020exponential,berdou2023one,ding2025quantum}. 
The previous topological QEC codes, however, are designed to correct any type of physical qubit error~\cite{PhysRevLett.84.2525}, providing equal protection against $X$ and $Z$  errors, and do not thus  exploit the noise bias. The possibility to exploit the physical asymmetry for a dominant dephasing noise to improve logical performance   has motivated the recent development of various QEC strategies.

In the context of topological QEC codes, a paradigmatic example of a bias-tailored code is the so-called XZZX code~\cite{PhysRevX.9.041031}, which is constructed from the Calderbank-Shor-Steane (CSS) rotated surface code~\cite{PhysRevA.76.012305,PhysRevA.90.062320} by applying Hadamard (H) gates to alternating physical qubits, thereby reshaping the stabilizer structure to better match the typical  errors occurring under an asymmetric biased noise (see Fig.~\ref{fig:surface-codes}b). Under an idealised code-capacity model biased towards dephasing $Z$ errors, which only acts on the physical qubits in between  rounds of ideal syndrome extraction, the XZZX code exhibits an error threshold that approaches $p_{\rm th}=0.5$ in the limit of pure dephasing. This saturates the classical Hashing bound, and provides a considerable improvement with respect  to the   code-capacity threshold of the conventional surface code under un-biased depolarising noise $p_{\rm th}=0.189$~\cite{PhysRevX.2.021004,PhysRevLett.109.160503}. This advantage is progressively reduced as the relative bias of the noise is decreased until, in the limit of  equally-likely phase- and bit-flip errors,   both codes provide a comparable performance~\cite{BonillaAtaides2021}. This  result suggested that one can substantially enhance logical performance by tailoring the code structure to the dominant physical error channel, motivating a  range of subsequent works on biased-tailored QEC. Some examples include finite-bias variants of the XZZX code optimizing effective-distance scaling~\cite{PhysRevResearch.5.013035}, or the extension of bias-tailoring approaches  beyond topological codes to  quantum low-density parity check~\cite{Roffe2023biastailoredquantum,6qbg-xslr,rrfk-1b51,wu2025biastailoredsingleshotquantumldpc} and Floquet~\cite{Setiawan2025} codes.   Altogether,  these works suggest that biased-tailored  QEC  is a promising route to accelerate the progression towards the teraquop regime but, at the same time,  raise the crucial question of whether such bias-induced advantages survive realistic gate-level errors and their circuit-level  propagation, both of which go beyond the idealised model of biased code-capacity noise.

This fundamental tension was already recognized in the early work of Aliferis and Preskill~\cite{aliferis2008fault}, who pointed out that the preservation of  dephasing bias  depends  on the microscopic structure of the native gate set. In particular, while diagonal entangling gates  such as the controlled-Z (CZ) gate naturally preserve this noise asymmetry, generic non-diagonal operations such as the controlled-NOT (CNOT)  can convert phase errors into bit-flip ones, partially degrading the resulting bias. In fact,  it is proven that one cannot implement all the required physical gates required for QEC in a way that fully preserves the noise bias when working with two-level  qubits~\cite{PhysRevX.9.041053}. Therefore, any realistic syndrome-extraction circuit will inevitably  reduce the effective noise asymmetry, and  understanding how much of the potential  advantage does survive is a central question for  practical bias-tailored QEC.
This question has recently re-emerged in the context of   topological codes, where it was shown  that a large fraction of the  performance advantage for the XZZX code is lost under realistic circuit-level noise~\cite{martinez2025leveragingbiasednoiseefficient}. Interestingly, some residual improvement still remains  for architectures where the transpilation of the CNOT operations into the native entangling gate set  retain a certain degree of asymmetry in the noise.

In this work, we provide further quantifications of the actual advantage of bias-tailored QEC strategies under real circuit-level noise. We start by addressing a natural question: is the XZZX code advantageous against a more naive solution in which the standard surface code can be scaled anisotropically by  adaptively changing the ratio of the logical distances? (see Fig.~\ref{fig:surface-codes}c). A priori,  this approach could be interesting for  architectures such as trapped-ion quantum charged coupled device (QCCD)~\cite{Kielpinski2002,Pino2021}, in which the  layout of the physical-qubit registers can be reconfigured by exploiting physical ion shuttling and reordering~\cite{PhysRevX.7.041061,PhysRevA.99.022330,PhysRevA.100.062307,benito2025scalingroadmapmodulartrappedion}.
Interestingly, we find that even if the anisotropy-optimised surface codes can in principle outperform the XZZX codes in a certain bias regime for code-capacity noise model, the XZZX code  displays a slightly better performance for arbitrary bias once realistic circuit-level noise is included. 

In the second part of our work, focusing on the proven superior XZZX code,  we address another question: can one design an FT strategy to reduce the bias degradation during the faulty syndrome extraction, such that the potential advantage of the XZZX code   under circuit-level noise is not degraded so much?. In particular,  we introduce an encoded  CNOT gadget   acting as a deterministic filter  that suppresses the propagation of un-biased errors, enhancing the bias preservation  relative to that of the bare CNOT gates that are used in  standard XZZX syndrome extraction circuits. This bias filter  works by temporarily encoding the target qubit in a repetition code using additional ancillary qubits and using feed-forward operations conditioned on the ancillary measurement results. We incorporate this bias-filtering  CNOT gadget into the syndrome-extraction circuit of the XZZX code, and identify the noise regimes in which it improves the QEC threshold relative to the standard implementation of the XZZX code based on bare CNOT gates. The gadget achieves this improvement by suppressing non-bias-preserving error processes at the cost of increasing the circuit complexity and qubit overhead, effectively leading to additional bias-preserving errors. This translates into a trade-off between enhanced bias conservation and accumulated fault-tolerant resources that only pays off for high bias and small idle errors compared to entangling-gate errors. 

This article is organized as follows. We start in Sec.~\ref{sec:noise-bias} by defining the biased noise models that we use in our simulations, and describing the effects of non-bias-preserving CNOT gates under circuit-level noise. We proceed in Sec.~\ref{sec:anisotropic-scaling} by comparing the performance of the XZZX code with a naive anisotropic scaling of the CSS surface code. Then, in Sec.~\ref{sec:gadget}, we introduce a bias-filtering gadget that implements a bias-preserving CNOT. We characterize the gadget and use it as a building block for the XZZX code. Finally, in Sec~\ref{sec:conclusions}, we summarize the results and discuss future directions.

\section{\bf Circuit-level Pauli noise models}\label{sec:noise-bias}
The standard circuit-level noise model of a noisy quantum computer assumes that each operation targeting a specific single- or two-qubit gate unitary, state preparation or readout, is modeled by the ideal quantum operation followed by an error channel that quantifies the   deviations from the ideal targets. Assuming that there are no temporal or spatial noise correlations between the different operations in a specific  circuit, the most-general error  allowed by physics is described by a fixed set of  completely-positive and trace-preserving  (CPTP) maps $\mathcal{E}$, each applied to subsets of the  $N$-qubit register. For a density matrix  $\rho$, these maps have the form 
\begin{equation}
\mathcal{E}(\rho)=\sum_iK_i\rho K_i^\dagger,
\end{equation}
where the Kraus operators $K_i\in\mathsf{L}(\mathcal{H})$ are specific linear operators with support on a subset of qubits within the $N$-qubit Hilbert space $\mathcal{H}=\mathbb{C}^{2N}$, and subject to the trace constraint $\sum_iK_i^\dagger K_i=I_N$~\cite{Nielsen_Chuang_2010}. Hence, when performing a circuit-level assessment of logical QEC performance, the ideal quantum operations in the FT circuits are exchanged for faulty ones, e.g. unitary gate $U_{\rm t}\rho U_{\rm t}^{\dagger}\mapsto \mathcal{E}_t(U_{\rm t}\rho U_{\rm t}^{\dagger})$ where the qubit support and structure of $\mathcal{E}_t$ depend on the specific operation, and the  specific noise rates are encoded in the Kraus operators. 

The assumptions of no spatial correlations set constraints on the support of these CPTP channels, as $m$-qubit gates only affect the specific set of $m$ qubits. For temporal correlations, even if those can be encoded in the effective form of error rates and Kraus operators $K_i$ during a single operation, one assumes that these correlations do not extend to subsequent gates in the circuit, such that the same CPTP map is always applied following  a specific type of gate in the circuit, regardless of any previous history of gates taking place earlier. Moreover,  for the purpose of benchmarking QEC codes at scale, CPTP maps   are further restricted  to be Pauli channels
\begin{equation}
\mathcal{E}_{\rm P}(\rho)=(1-p)\rho+\sum_{i=1}^{4^N-1}p_iP_i\rho P_i,
\label{eq:pauliop}
\end{equation}
where $p=\sum_ip_i\in[0,1]$, and the Kraus operators belong to the $N$-qubit Pauli basis $P_i\in \mathcal{P}_N\equiv\{I,X,Y,Z\}^{\otimes N}\setminus I_N$, where $X,Y,Z$ are the single-qubit Pauli matrices. This structure can be interpreted as the random occurrence of each of the Pauli operators $\{P_i\}$  according to the  probability distribution $\{p_i\}$. Since QEC syndrome-extraction circuits usually belong to the Clifford group, mapping thus Pauli operators onto Pauli operators,  restricting the noise to Pauli channels considerably simplifies their numerical simulations~\cite{PhysRevA.70.052328}, and is currently performed by efficient algorithms that perform Pauli-frame updates~\cite{Gidney2021stimfaststabilizer}. Additionally, using  Pauli noise enables the automated generation of the detector error model (DEM), which tells the decoder how error events violate the parity measurements of the code's error syndrome, and is very practical for improved decoding. 

The most common Pauli channel used in the context of QEC is the depolarising error model, which sets equal weights to all possible Pauli errors acting on $m$ qubits $p_i=p_{\rm dep}/(4^m-1)$, such that the depolarising error rate $p_{\rm dep}\in[0,1]$ is the probability with which any of the Pauli errors can occur
\begin{equation}
\label{eq:dep_channel}
\mathcal{E}_{\rm dep}(\rho)=(1-p_{\rm dep})\rho+\frac{p_{\rm dep}}{4^m-1}\sum_{i}P_i\rho P_i.
\end{equation}
This channel effectively admixes the quantum state $\rho$ with the maximally mixed state $I_m$ without selecting any preferred axis. However, as advanced in the introduction, it is common for several experimental platforms to have a preferred   axis, as in the case of dominant dephasing noise.
Generally, a biased-dephasing  Pauli channel will have higher error probabilities $p_i$ for those operators $P_i\in\mathcal{P}_N$ that only contains $Z$ operators $
\mathcal{P}_N^\text{bias}= \{I,Z\}^{\otimes N}\cap\mathcal{P}_N$. To quantify the amount of bias, we classify the error components of the channel depending on whether they have non-$Z$ operations acting on any of the qubits, such that the corresponding biased and unbiased error rates    read 
\beq
\label{eq:bias_rates}
p_\text{bias}=\sum_{\{i\,:\, P_i\in \mathcal{P}_N^\text{bias} \}}\!p_i,\hspace{2ex
}p_\text{unbias}=\sum_{\{i\,:\, P_i\notin \mathcal{P}_N^\text{bias} \}}\!p_i.
\eeq

Using this notation, the relative bias  of a noise channel 
\beq
\eta=\frac{p_\text{bias}}{p_\text{unbias}}
\eeq
quantifies  the probability  ratio between  errors occurring along the preferred axis or along any of the other axes.
A Pauli error channel is thus biased when the average probability of biased components is greater than that of unbiased ones, which happens for 
\beq\eta>\frac{\lvert \mathcal{P}_N^\text{bias}\rvert}{ \lvert\mathcal{P}_N\setminus\mathcal{P}_N^\text{bias}\rvert}=\frac{1}{2^N}.
\eeq
An operation is said to be bias-preserving if its associated noise channel has a similar bias with respect to any other operation in the quantum system. The aforementioned code-capacity and phenomenological biased error models fall in this category, as they only consider the effect of  the biased error channel  on qubits in between perfect syndrome extraction, or model instead readout imperfections by a phenomenological fault during syndrome extraction that does not propagate along the circuit nor does it alter the bias. In general, however, error propagation in a circuit need not respect the relative bias, and must be carefully accounted for a realistic assessment of the practical advantage of biased-tailored codes.

\subsection{Standard depolarising  noise model}
The most common  circuit-level error model in studies of QEC performance uses the depolarizing channel~\eqref{eq:dep_channel} for single-qubit H gates and idle periods and two-qubit CNOT gates $CNOT=(I\otimes I+Z\otimes I+I\otimes X- Z\otimes X)/{2}$, and a bit-flip channel for  state preparation/reset and measurements. Typically, one sets all the error rates of these channels to be equal, and refers to this model as the standard or uniform depolarising error model.

\subsection{Biased dephasing   noise model}\label{sec:biasedcircnoise}
To compare quantitatively with previous accounts of bias-tailored codes in the presence of circuit-level noise~\cite{martinez2025leveragingbiasednoiseefficient},
 we consider instead a multi-parameter biased Pauli   error model in which
\begin{itemize}
\item Hadamard  gates $H=(X+Z)/\sqrt{2}$ can fail according to a fully depolarizing channel with rate $p_\text{1q}$.
\item Controlled-Z  gates $CZ=(I\otimes I+Z\otimes I+I\otimes Z- Z\otimes Z)/{2}$ can fail according to a uniformly-biased two-qubit Pauli channel~\eqref{eq:pauliop} with rate $p_\text{2q}$ and bias $\eta_\text{2q}$. Biased components $\{I\otimes Z,Z\otimes I,Z\otimes Z\}$ happen with probability $p_{{\rm bias},i}={p_\text{2q}\eta_\text{2q}}/{3(\eta_\text{2q}+1)}$ and the rest with $p_{{\rm unbias},i}={p_\text{2q}}/{12(\eta_\text{2q}+1)}$.
\item Projective measurements $M_{o}=\ket{o}\!\bra{o}$  fail by flipping the outcome $o\in\{0,1\}$ with probability $p_\text{m}$.
\item Resets with Kraus operators $R_0=\ket{0}\!\bra{0},R_1=\ket{0}\!\bra{1}$ can fail when preparing a $\ket{o}$ state, by preparing instead $\ket{1-o}$  with probability $p_\text{r}$.
\item Identity evolution $I$ on idle qubits can fail due  to environmental errors modeled by a biased  Pauli channel~\eqref{eq:pauliop} with a rate $p_\text{id}(t)=q_\text{id}t$ that depends on the idle time $t$,  and a relative  bias $\eta_\text{id}$. The $Z$ error happens with probability $p_{\rm bias}={p_\text{id}(t)\eta_\text{id}}/({\eta_\text{id}+1})$ while $X$ and $Y$ occur with probability $p_{{\rm unbias},x}=p_{{\rm unbias},y}={p_\text{id}(t)}/{2(\eta_\text{id}+1)}$.
\end{itemize}

\begin{figure}
\includegraphics[scale=0.6]{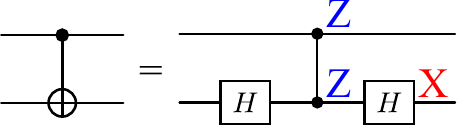}
\caption{{\bf CNOT to CZ transpilation:} standard transpilation of a CNOT gate into a controlled-Z gate and two Hadamards. Even if the CZ gate is constructed to ensure $Z$-noise bias, the H gates transform it to $Z\otimes X$, losing the bias structure. The H gates introduce additional non-biased noise to the target qubit.}
\label{fig:transpile_cz}
\end{figure}

In Ref.~\cite{PhysRevX.9.041053}, it was shown that a CNOT gate cannot be implemented as a fully bias-preserving operation when working with two-level qubits, in contrast to diagonal entangling gates such as the CZ gate, which naturally preserve the bias. A clear example is the straightforward transpilation of a CNOT into a CZ gate surrounded by Hadamards acting on the target qubit, as illustrated in Fig.~\ref{fig:transpile_cz}. Even if the underlying CZ gate is  bias preserving, the Hadamard rotations transform phase-flip errors into bit-flip errors according to $ZH=HX$, thereby degrading the relative bias. Consequently, the syndrome-extraction circuits for CSS topological codes, which require sequences of CNOT operations, inevitably degrade the effective noise asymmetry, which can substantially reduce the performance gains predicted for bias-tailored codes under idealized code-capacity noise models. This motivates the careful assessment of bias-tailored QEC  under biased circuit-level noise.

\section{\textbf{Bias-tailored QEC:}  XZZX  codes versus  optimised  rectangular surface codes}\label{sec:anisotropic-scaling}

We start by performing a detailed comparison of the bias-tailored XZZX code with a more direct bias-tailoring strategy that can be exploited for some CSS codes such as hypergraph product codes~\cite{6671468}, considering different code distances for the $X$ and $Z$ logical operators.
In the surface code, this can be achieved by using rectangular lattices, elongating  the rectangle along the dominant error direction to find better logical performance with respect to the isotropic surface code on a square lattice. However,  these anisotropic surface-code approaches were typically analyzed under the idealized code-capacity biased-noise models~\cite{lee2021rectangular}, and no explicit quantitative  comparison to the XZZX code was performed.   Moreover, the performance of this anisotropic approach under faulty syndrome-extraction circuits and gate-level error propagation has, to the best of our knowledge, not been previously addressed in detail, leaving open the question of how their relative performance behaves in realistic circuit-level descriptions. To perform this comparison, let us start by defining the QEC codes. 

The surface code is a topological stabilizer QEC code defined on a two-dimensional square lattice with qubits placed on the links~\cite{bravyi1998quantumcodeslatticeboundary,10.1063/1.1499754}. In its rotated version~\cite{PhysRevA.76.012305,PhysRevA.90.062320}, its stabilizer group is generated by adjacent  weight-4  plaquette operators corresponding to products of Pauli-$X$ and Pauli-$Z$ operators $XXXX$ and $ZZZZ$, with smaller-weight stabilizers appearing at the boundaries (see Fig.~\ref{fig:surface-codes}a). For biased noise,  the qubits can be arranged on  an anisotropic rectangle, and its sides can be independently scaled. Since logical $X_L$ and  $Z_L$ operators correspond to non-contractible paths connecting opposite boundaries of the lattice, this anisotropic scaling allows the associated logical distances $d_Z$ and $d_X$ 
to be independently adjusted (see Fig.~\ref{fig:surface-codes}c). Under dominant dephasing noise, one increases $d_Z$ to
enhance thr protection against physical $Z$-error chains built from the occurrence of phase-flip errors on contiguous qubits, which could potentially lead to an eventual logical $Z$ failure.

 The XZZX code is a bias-tailored variant of the square-lattice surface code that maintains its $\mathbb{Z}_4$ rotational symmetry, and instead   applies Hadamards to alternating qubits~\cite{BonillaAtaides2021}. Under this transformation, the conventional $XXXX$ and $ZZZZ$ stabilizers are mapped into $XZZX$-type stabilizers, while retaining their mutual commutativity (see Fig.~\ref{fig:surface-codes}b). In contrast to the CSS surface code, where $X$- and $Z$-errors are detected independently, each mixed stabilizer now simultaneously probes both types of errors. Under dominant dephasing noise, this mixed stabilizer structure effectively aligns the dominant physical $Z$-error chains along a preferred direction of the decoding graph in the DEM, causing the resulting error-correction problem to behave similarly to that of a repetition code. In fact, the code exhibits a substantially enhanced robustness against biased dephasing noise, achieving a code-capacity threshold that approaches $p_{\rm th}=0.5$ in the limit of  pure dephasing. In this section, we compare the performance of the XZZX code against the previous rectangular  surface codes with the same number of physical qubits.

\begin{figure}
\centering
\subfloat[]{\includegraphics[width=.9\linewidth]{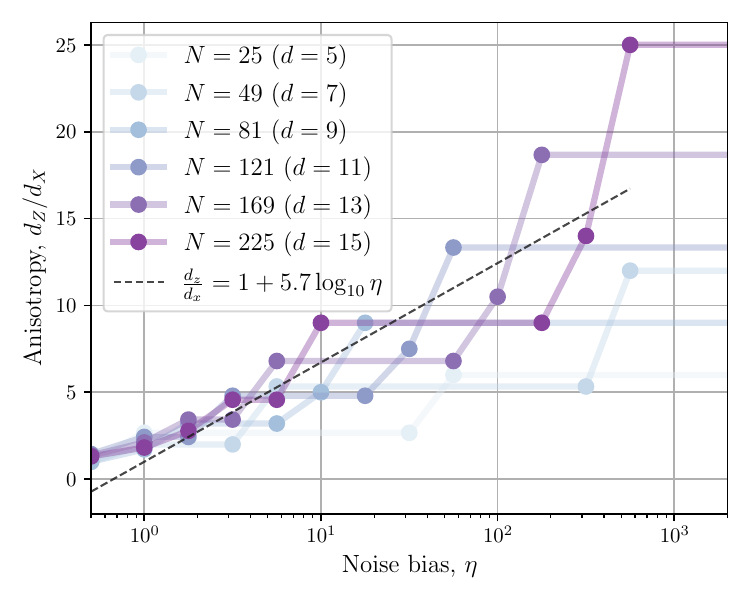}}\\
\subfloat[]{\includegraphics[width=.95\linewidth]{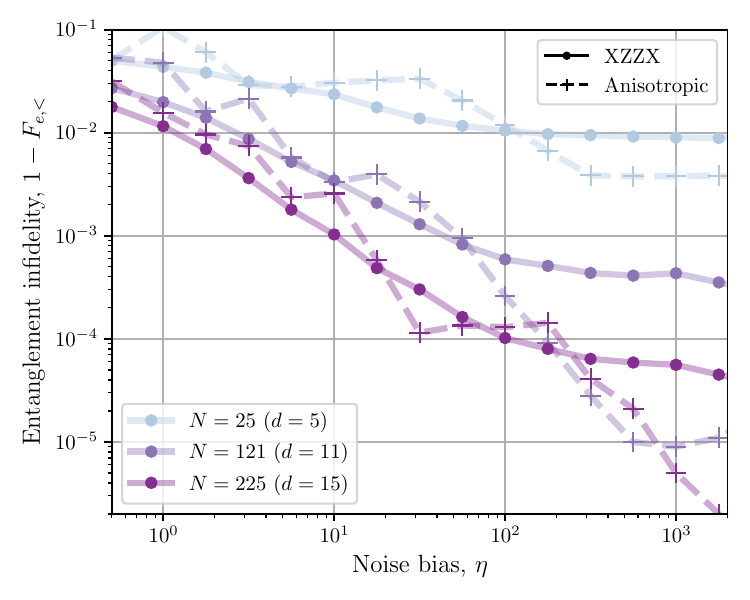}}
\caption{{\bf XZZX and anisotropic surface code under biased code-capacity noise} 
a) Optimal anisotropy that maximizes the entanglement fidelity of the logical qubit, as a function of the bias. The maximum anisotropy for a given distance is limited by the distance itself, which corresponds to having a repetition code instead of a surface code. In this work, we do not consider repetition codes as we limit the minimum distance to $d_X=2$. We perform a linear fit of the anisotropy to $1+m\log\eta$, considering only points from the $N=225$ curve with $d_X>3$ to limit finite-size effects, that we plot as a dashed grey line. b) Comparison of the logical entanglement infidelities between the XZZX (solid dots) and optimal-anisotropy  surface codes (dashed stars) for biased code-capacity noise with rate $p=0.1$ as a function of bias. For smaller bias, the XZZX code exhibits a better protection for quantum memories, generally outperforming the optimal-anisotropy rectangular surface codes. However, for higher biases, anisotropic codes effectively get closer to repetition codes, which have a linear (instead of quadratic) dependence between the number of qubits and the code distance. We note that the pikes in the performance of the anisotropic codes are associated to changes in the optimal anisotropy, which shows a stepped behavior due to code distances being discrete numbers.
}
\label{fig:anisotropic_vs_bias}
\end{figure}

As a warm up, we start this comparison by benchmarking both approaches under   biased code-capacity noise acting on the physical qubits in between   perfect syndrome extraction rounds (see Fig.~\ref{fig:anisotropic_vs_bias}).
For this plot, we ideally prepare the logical $\ket{0}_L$/$\ket{+}_L$ states, we apply a biased noise channel to all data qubits, and perform ideal syndrome extraction. Finally, we perform an ideal logical measurement in the $Z$/$X$ basis, which we correct using the PyMatching decoder~\cite{Higgott2025sparseblossom} that runs on numerically-simulated syndrome-extraction data. By repeating the experiment multiple times, we can estimate the logical failure probabilities $p_0$ and $p_+$. 

The XZZX code mixes $X$ and $Z$ errors and  the logical error rates turn out to be the same $p_0=p_+$. In contrast, as $X$ and $Z$ errors are detected independently in the rectangular surface code,  the performance can be different for a distinct input state due to the bias. Therefore, to learn how the lattice anisotropy has to be optimized for a certain bias, we need a metric that takes into account both logical errors on the same footing.
We use the entanglement fidelity  $F_e$~\cite{PhysRevA.54.2614,NIELSEN2002249} as  a first-principles metric in that regard,  which quantifies how well a noisy logical channel preserves the full quantum information encoded by the specific QEC code in an arbitrary logical input state.
 For a logical qubit evolving under a logical Markovian error channel, the entanglement fidelity should in principle also include the  $Y_L$ basis, as it can be reconstructed from the logical error probabilities associated with the six cardinal states $\{\ket{0}_L,\ket{1}_L,\ket{+}_L,\ket{-}_L,\ket{+i}_L,\ket{-i}_L\}$.
For the surface code, however, encoding of $\ket{+{\rm i}}_L$ states and $Y$-type measurements are not easily implemented, as not the whole logical Clifford group is transversal.  We thus use a strict lower bound on the entanglement fidelity~\cite{vezvaee2025surfacecodescalingheavyhex} that reads as follows
\begin{equation}
\label{eq:EnInf}
F_{e,L}=\bra{\phi} \mathcal{I}\otimes\Psi\left(\ket{\phi}\bra{\phi}\right)\ket{\phi}\geq F_{e,L}^{<}=(1-p_0)(1-p_+),
\end{equation}
where $\Psi$ is the    quantum channel for the logical QEC memory, $\mathcal{I}$ is the identity channel, and
$\ket{\phi}=(\ket{0_L,0_L}+\ket{1_L,1_L})/\sqrt{2} $ is a maximally-entangled logical state.

Under the assumption of independent logical bit- and phase-flip errors, this lower bound is actually exact $F_{e,L}=F_{e,L}^{<}$~\cite{vezvaee2025surfacecodescalingheavyhex}.
For the anisotropic surface code, we depict in Fig.~\ref{fig:anisotropic_vs_bias}a) the optimal code anisotropy ratio $d_Z/d_X$ minimizing the entanglement infidelity bounds, and therefore maximizing the average logical performance of the quantum memory. One observes how the optimal layout of the code  for isotropic distances up to $d=15$ with a fixed number of qubits increases in steps as the noise bias grows. This staircase behavior reflects the fact that, at a certain bias level,  it becomes advantageous to increase (decrease) $d_Z$ ($d_X$) distance such that one gains an increased protection against phase flips at the expense of a smaller protection against bit flips. It is interesting that the staircase has smaller step widths as the total qubit number increases, and the anisotropy scales logarithmically with respect to bias, up to finite-size effects. An expression that approximates the optimal anisotropy as a linear function of $\log\eta$ is derived in App.~\ref{app:optimal-anisotropy}, which captures quantitatively the anisotropic scaling with noise bias as shown in a dashed line in Fig.~\ref{fig:anisotropic_vs_bias}{a)}.

In Fig.~\ref{fig:anisotropic_vs_bias}b), we present numerical results for  the entanglement infidelity bounds of such an optimal-anisotropy rectangular surface code (stars), and compare it to corresponding entanglement fidelity of the XZZX quantum memory (circles)  under biased code-capacity noise. We  observe that the XZZX code is  advantageous for small and intermediate bias, but the optimal-anisotropy rectangular surface code eventually becomes the best performing approach, which is marked by a crossing of the respective curves  in the high-bias regime. This behavior can be understood for the asymptotic behavior of both codes in the limit of infinite bias. Here,  the anisotropic surface code becomes a repetition code with distance $d=N$, whereas the XZZX code only achieves $d=\sqrt{N}$ scaling. Consequently, although both approaches exhibit the same asymptotic threshold under fully dephasing noise, the anisotropic construction can achieve a larger effective distance and therefore a lower logical error rate at fixed QEC footprint as quantified by the $F_{e,L}$.

As already discussed in the introduction, logical performance estimates based on code-capacity noise can be deceptive, as they tend to overestimate the practical advantages of bias-tailored QEC. In general, we expect that realistic syndrome-extraction circuits will  reduce the effective noise asymmetry through non-bias-preserving operations and gate-level error propagation. Here, we show that code-capacity studies can even lead to qualitatively misleading conclusions, namely predicting a large-bias regime in which anisotropic surface codes outperform the XZZX code.
We quantify the performance under circuit-level noise by simulating a quantum memory experiment where we apply $d$ rounds of noisy syndrome extraction. The first  round is used to prepare the logical qubit in the desired state, as this is a valid FT initialization procedure for CSS codes. Since the XZZX code is equivalent to the CSS surface code up to single-qubit unitaries, the same protocol is also valid to prepare logical $\ket{0}_L$ and $\ket{+}_L$ states in the XZZX code. After $d$ syndrome extraction cycles, we measure all data qubits to perform a logical measurement. The final measurement provides additional syndrome information to ensure fault-tolerance of the measured logical state. The syndrome information from all cycles is provided to the PyMatching decoder, which indicates whether a correction has to be applied to the measurement outcome of the logical operator. Finally, by repeating the simulation multiple times, we obtain $p_0$ and $p_+$ to bound the entanglement infidelity~\eqref{eq:EnInf}.
 
\begin{figure}
\centering
\includegraphics[width=\linewidth]{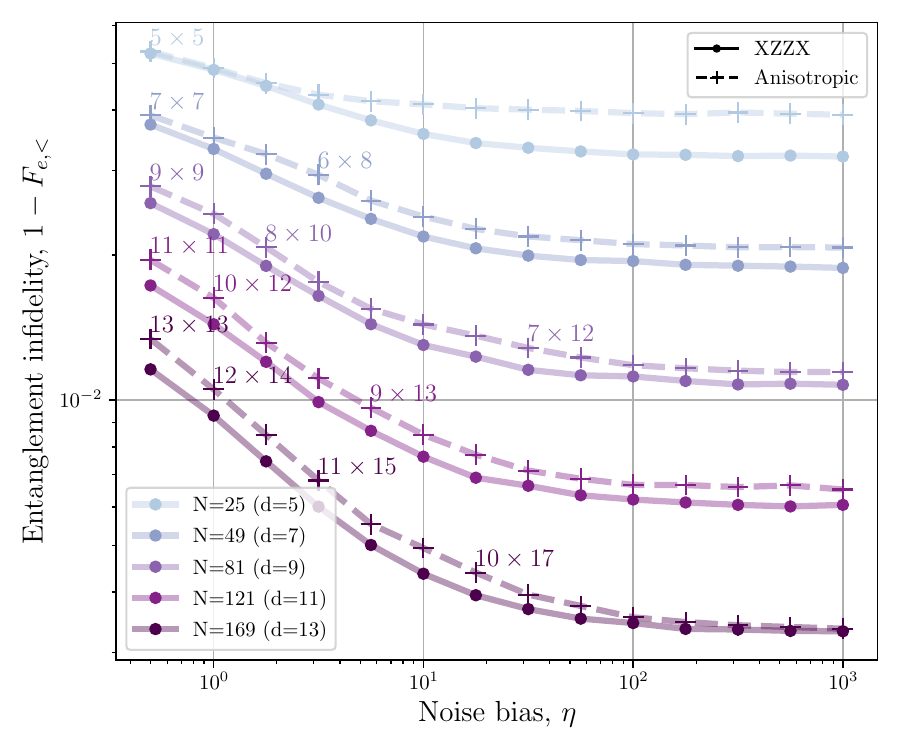}
\caption{{\bf XZZX and anisotropic surface code under biased circuit-level noise}: infidelity comparison between the XZZX (solid dots) and anisotropic (dashed stars) surface code under circuit-level noise with $p=0.003$ and variable bias for idle qubits and two-qubit gates. For both codes, there is a residual improvement on the logical error rates under biased noise, but it saturates at $\eta\sim100$ due to the CNOT gates used in syndrome extraction producing non-biased noise. For higher distances, where more anisotropy levels are available, anisotropic surface codes match the performance of the XZZX code.
}
\label{fig:anisotropic_vs_bias_circuit}
\end{figure}

 In Fig.~\ref{fig:anisotropic_vs_bias_circuit}, we compare the performance of the anisotropic surface code  with optimised anisotropic ratios and the XZZX code, both under  biased circuit-level noise, considering distinct bias values for idling qubits and CZ gates.
 As  shown in the figure, both approaches display a quantitatively-similar logical performance, which shows that the apparent large-bias advantage of the anisotropic strategy vs the XZZX code  is actually an artifact of the strong  assumptions underlying the code-capacity noise model. Moreover, the substantial improvements in logical entanglement fidelity predicted under code-capacity noise as the bias increases, which were predicted to span several orders of magnitude for large code distances in Fig.~\ref{fig:anisotropic_vs_bias}, become dramatically reduced once realistic circuit-level effects are included.

These results suggest that, under realistic circuit-level noise, the dominant limitation of biased-tailored QEC is not necessarily the specific stabilizer structure of the code itself, but rather the degradation of the physical noise asymmetry during syndrome extraction. Motivated by this observation, in the next section we investigate how certain  bias-filtering circuit constructions can recover part of the lost advantage. We  introduce an encoded CNOT gadget designed to reduce bias degradation by noise filtering during syndrome extraction, which we incorporate into the XZZX code owing to its slightly superior performance with respect to rectangular surface codes under circuit-level biased noise.

\section{\bf Bias-filtering  CNOT gadget}\label{sec:gadget}
\begin{figure*}
\centering
\subfloat[\label{subfig:gadget}]{\includegraphics[width=.5\linewidth]{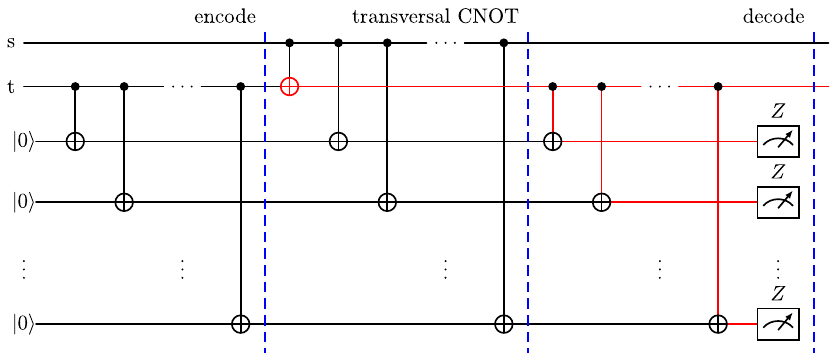}}
\subfloat[\label{subfig:gadget-cz}]{\raisebox{.75cm}{\includegraphics[width=.5\linewidth]{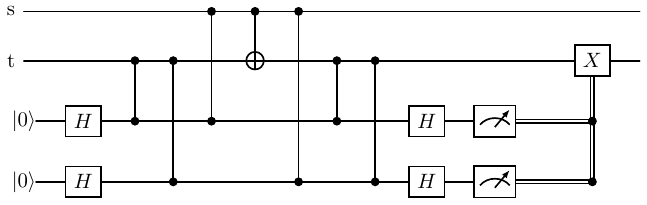}}}
\caption{{\bf Bias-filtering CNOT gadget:} {\bf a)} Repetition-code implementation of the bias-filtering gadget. It implements a CNOT gate by encoding the target qubit in a distance $d$ repetition code, performing a transversal CNOT, and finally decoding from parity measurements. A bit flip is applied to the target qubit if more than $d/2$ measurements are flipped. This allows detecting and correcting the non-biased bit-flip originated in the original CNOT between the source and target qubits, displayed in red. {\bf b)} Simplified gadget for the distance-3 repetition code, using a depth-5 circuit. The CNOT gates are simplified to CZ gates by canceling out Hadamard gates under the assumption that gate errors dominate over idling errors.}
\label{fig:gadget}
\end{figure*}

As already remarked, a central challenge for biased-noise QEC is that syndrome-extraction circuits inevitably contain operations that partially convert dominant dephasing errors into unbiased ones, progressively reducing the effective noise asymmetry and the corresponding bias-tailoring advantage. Refs.~\cite{PhysRevA.78.052331,PhysRevA.92.062309} addressed this problem by adapting the entire QEC architecture to the underlying noise asymmetry. Embedding repetition-code layers together with native diagonal entangling gates within concatenated CSS-code constructions, these works showed that the logical performance can be improved under strongly biased noise. However, as a direct consequence of the concatenation, the resulting FT
 constructions effectively require non-local logical operations and increasingly complex syndrome-extraction procedures, moving away from the useful local connectivity and sparse syndrome structure characteristic of topological codes. Moreover, these schemes cannot naturally exploit the highly optimized decoding strategies available for topological QEC such as anisotropic surface and XZZX codes.

We circumvent these limitations by devising a  bias-filtering CNOT gadget that  encodes the target qubit into a repetition code only temporarily, and subsequently decodes it while applying feed-forward operations conditioned on ancillary-qubit  readout outcomes. In this way, the gadget acts as a local asymmetry filter that selectively suppresses bias-degrading error propagation at the circuit level, while respecting the native local structure, sparse decoding graphs, and practical scalability advantages of the topological XZZX code.
In particular,  our bias-filtering gadget encodes the target qubit  into a quantum $d$-qubit repetition code using a sequence of CNOT gates to $(d-1)$ ancilla qubits (see Fig.~\ref{fig:gadget}).
The original syndrome-extraction CNOT gate between the source physical qubit and the encoded target qubit is then implemented by applying pair-wise CNOTs by exploiting transversality. After the encoded CNOT operation, the encoded target qubit is  decoded by applying the same sequence  of CNOTs. To enable the bias filtering,  the $(d-1)$ ancillary qubits are subsequently measured to obtain the repetition-code parities, which contain information about   undesired bit-flip errors that may have propagated through the physical CNOT gates degrading  the relative bias. Since  only  errors on the target physical qubit are of relevance to the bias preservation (see Fig.~\ref{subfig:gadget}), the correction must only apply an $X$ correction to the decoded target qubit when at least $(d+1)/2$ measurements of the repetition code ancillary qubits turn out to be flipped. This filtering strategy can be seen as a flag-qubit-type gadget~\cite{PhysRevLett.121.050502,Chao2018,Chamberland2018flagfaulttolerant}, albeit it does not filter out error propagations that compromise FT, but instead a portion of those that do not respect the noise bias.

Let us note that the circuits in Fig.~\ref{fig:gadget} are only FT against the biased $X$ errors that originate in the target qubit of all CNOT gates. However, there are other locations which, although suppressed by a factor of $1/\eta$, introduce uncorrectable $X$ errors that can appear in the original target qubit, as is the case of errors from the CNOTs in the encoding layer. Thus, $X$ errors cannot be arbitrarily suppressed below what bias-preserving gates introduce, and the effective noise channel of the CNOT gadget is limited by
\begin{equation}
\begin{split}
    p_{X,Y}^s&=\mathcal{O}(\frac{dp}{\eta}),\,\, \hspace{9ex}p_Z^s=\mathcal{O}(dp),\\
    p_{X,Y}^t &= \mathcal{O}\left(\frac{dp}{\eta}+p^\frac{d+1}{2}\right),\,\,
    p_Z^t=\mathcal{O}(dp).\\
\end{split}\label{eq:gadgetrate}
\end{equation}
We restrict ourselves to repetition codes of $d=3$, as higher distances would increase $Z$ errors without a substantial reduction on $X$ errors due to the error floor present in $p_{X,Y}^t$ from Eq.~\eqref{eq:gadgetrate}, which prevents arbitrary suppression of non-bias-preserving errors below $\mathcal{O}({dp}/{\eta})$.

\begin{figure*}
    \centering
    \includegraphics[width=\linewidth]{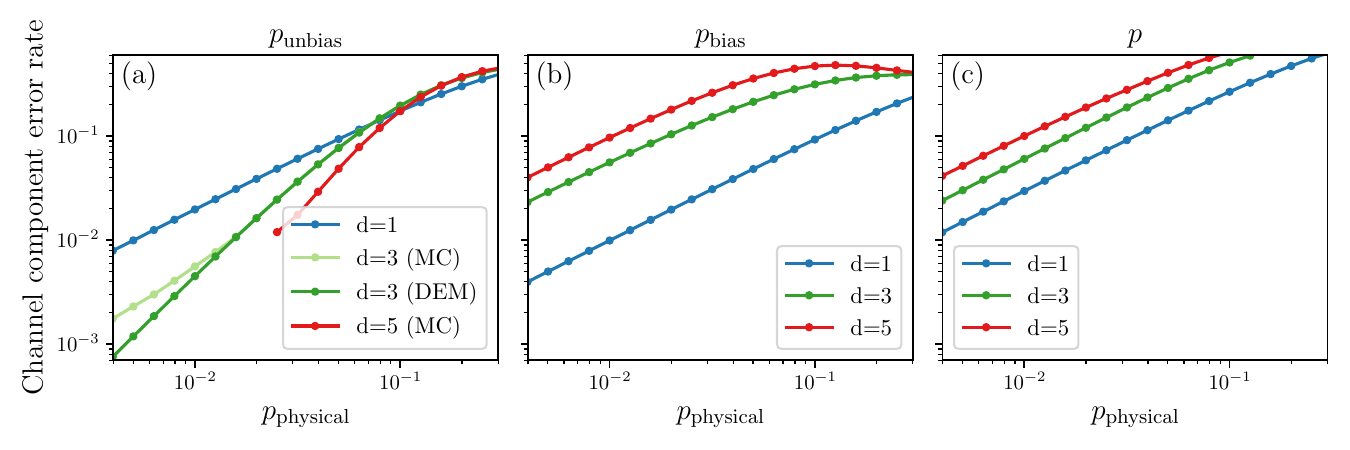}
    \caption{{\bf Performance of bias-filtering CNOT}:  Unbiased (a), biased (b) and total (c) components of the effective error channel of the bias-filtering CNOT gadget. Represented as a function of the noise rate of physical gates, for different sizes of the inner repetition code. $d=1$ corresponds to the bare CNOT gate described in Fig.~\ref{fig:transpile_cz}. In (a), we show the differences between Monte-Carlo and DEM-based process tomography for the $d=3$ gadget. We observe that Monte-Carlo limits the maximum bias suppression observed by tomography. The red curve does not extend to all physical error rates, as the unbiased components for the $d=5$ gadget cannot be reliable computed from Monte-Carlo process tomography for such low  error rates, and the DEM approach is not available for $d=5$.
    }
    \label{fig:gadget-performance}
\end{figure*}

To further simplify the circuit, CNOT gates in the protocol are transpiled into CZ gates, and most $H$ gates cancel out. The resulting circuit, displayed in Fig.~\ref{subfig:gadget-cz}, has low overhead and thus can compete with a non-bias-preserving un-encoded CNOT. We now compare the performance of the bias-filtering  gadget to that of the bare CNOT gate, performing a full characterization of the encoded gadget using a simplified version of  quantum process tomography (QPT)~\cite{Nielsen_Chuang_2010}.

\subsection{Pauli channel tomography of the   gadget }\label{sec:characterization}
The bias-filtering CNOT gadget just introduced can be applied to multiple QEC circuits. To simulate their logical performance under biased noise, instead of performing circuit-level simulations including  the additional ancillary resources of the bias-filtering gadget and  the mid-circuit readouts,  we first characterize the effective error channel of the gadget using QPT (see Appendix~\ref{app:tomography}). Equipped with this error channel, we can efficiently simulate the  circuits for a bias-filtered XZZX quantum memory under our circuit-level noise model by inserting the obtained two-qubit error channel after each CNOT gate that appears during the syndrome readout. To characterize the noise in the bias-filtering gadget, we numerically perform process tomography to a circuit that implements the ideal CNOT gate followed by the noisy gates required to implement the bias filter. In this way, this should lead to an identity channel in absence of noise, and we can characterise the noise by deviations from this situation. 

Since the biased circuit-level  noise model defined in Section~\ref{sec:biasedcircnoise} only contains Pauli channels and the gadget  only contains Clifford operations, the effective error model  will also have the form of a Pauli channel~\eqref{eq:pauliop}. As explained in Appendix~\ref{app:tomography}, this simplifies the requirements of tomography: the qubits need only be initialized in all possible tensor products of the  $+1$ eigenstates of their respective  $X$, $Y$ and $Z$ Pauli operators, and only measured in the same respective basis $\boldsymbol{b}\in\{xx,xy,xz,yx,yy,yz,zx,zy,zz\}$. In this case, the probabilities $q_\mu$ are estimated using  the repeated outcomes  on all the required basis can be related to the Pauli channel error probabilities in~\eqref{eq:pauliop} by the linear map
 \beq
 q_\mu=\sum_{i=0}^{4^m-1}p_i\lvert\bra{\mu}P_i\ket{b,(+1)^{\otimes m}}\rvert^2,
 \eeq
 where $\ket{\mu}=\ket{\boldsymbol{b},\boldsymbol{m_b}}=\ket{b_1,m_{b_1}}\otimes\cdots\otimes\ket{b_N,m_{b_N}}$ denotes the state after measuring each qubit in the $\boldsymbol{b}$ basis and obtaining outcomes $\boldsymbol{m_b}$, $m=2$ is the number of qubits of the operation being characterized, and we defined $p_0\equiv1-\sum_{i=1}^{4^m-1}p_i$ for convenience. Then, the error channel can be determined by a least-squares inversion with a considerable reduction in the estimation complexity.

The most straightforward way to numerically simulate the measurement outcome distribution of the tomography circuits is to perform noisy Monte-Carlo simulations using Stim~\cite{Gidney2021stimfaststabilizer}. Then, for the bias-filtering gadget, we apply a majority vote decoder that applies a Pauli frame update to the target qubit if more than half of the parity checks are violated, correcting the obtained measurement outcomes. Finally, the probability is estimated from the counts of each outcome
\beq
q_\mu\approx\frac{N_{\boldsymbol{b},\boldsymbol{m_b}}}{N_{\boldsymbol{b}}}=\frac{N_{\boldsymbol{b},\boldsymbol{m_b}}}{\sum_{\boldsymbol{m_b}}N_{\boldsymbol{b},\boldsymbol{m_b}}}
\eeq
where $N_{\boldsymbol{b},\boldsymbol{m_b}}$ is the number of times that outcome $\boldsymbol{m_b}$ appears when measuring in the $\boldsymbol{b}$ basis, and $N_{\boldsymbol{b}}$ is the total number of shots for each basis.

In Fig.~\ref{fig:gadget-performance} we show the effective noise channel of the bias-filtering CNOT gadget, indicating the estimated Pauli error rates of biased and unbiased components~\eqref{eq:bias_rates} for multiple distances of the ancillary repetition block. The physical gates that are used to build this filter are simulated using the biased circuit-level noise model from Sec.~\ref{sec:biasedcircnoise}, but focusing on a single-parameter  where we set the bias $\eta_\text{2q}=1000$, $p_\text{physical}=p_\text{1q}=p_\text{2q}=p_\text{m}$ and $p_\text{id}=0$ to ease the discussion. The above MC  method has the problem that, for channels with high bias, some components are several orders of magnitude smaller than the rest, requiring a significantly larger number of shots to accurately estimate the non-biased error probabilities.
For sufficiently small circuits, we propose an alternative method to directly obtain the measurement outcome probabilities using the detector error models (DEMs) of the tomography circuits, as computed by Stim. The DEM exhaustively lists all independent error channels in the circuit, as well as the measurement outcomes that are flipped by them. From the DEM, we obtain all possible error configurations of the circuit, allowing us to determine the outcome distribution without approximate sampling, as described in Appendix~\ref{app:qpt-dem}.

Using our protocol, there is no precision loss due to Monte-Carlo sampling, but w remark that is only efficient for the $d=3$ gadget, as the number of error configurations in the DEM scales as $2^{2^{d+1}}$. For the current goal of this paper, as we are only considering applications of the bias-filtering gadgets with $d=3$,  the DEM protocol can be perfectly used without running into prohibitive exponential scaling bottlenecks. In Fig.~\ref{fig:gadget-performance}a), we compare the QPT of the bias-filtering gadget using the DEM and Monte-Carlo sampling approaches, which shows that the Monte-Carlo sampling  cannot fully capture the suppression of non-biased errors achieved by the bias-filtering CNOT in the small error regime. The suppression in this regime well captured by the DEM reconstruction, which shows a sustained slope for ever smaller error rates.

As the repetition block distance is increased, we observe in Fig.~\ref{fig:gadget-performance}a) a growing  suppression of the unbiased error components in the effective CNOT gate when the physical error rates are below a certain value, signaling the promised effect of the bias filtering. In Fig.~\ref{fig:gadget-performance}b),   we clearly observe the trade-off, as the biased error components increases with the code distance as a consequence of the higher circuit complexity. In Fig.~\ref{fig:gadget-performance}c), one can see how the overall error rate    is increased with respect to the bare ($d=1$) CNOT. 

We need to explore now if the advantage of the XZZX code stemming from the reduction of unbiased errors by  the bias-filtering gadget is sufficient when considering the overall increase of the biased errors. Once the gadget has been discussed and characterized, we add the bias-filtered CNOT to the available gate set, and its estimated Pauli error channel to the circuit-level noise model  used in the following  simulations, such that no extra numerical overhead appears in the  XZZX circuits. In this sense, we simply need to replace the error channel in the noise model that is applied in the simulation.

\subsection{Bias-filtered XZZX quantum memory}
The syndrome extraction circuit for the XZZX code requires two CNOT gates per stabilizer but, as  mentioned in previous sections, this pair of gates cannot be implemented  in a bias-preserving manner. Thus, syndrome extraction circuits introduce non-biased noise to the data qubits, limiting the advantage of the XZZX code. We show in Appendix~\ref{app:xzzx-circ} that, if all gates were bias preserving with the same physical error rate, the circuit-level threshold for the XZZX code would improve by a factor of 3 under biased noise. However, without a bias-preserving CNOT, the improvement on code threshold turns out to be only a 1.5 factor~\cite{martinez2025leveragingbiasednoiseefficient}. 

We try to further improve this logical performance  by using our bias-filtering CNOT gadget during syndrome extraction. The gadget introduces a tradeoff between the increased code threshold due to the bias-preserving CNOT and the higher error rates of the biased components of the CNOT noise channel. To find the regime where the CNOT gadget shows an advantage, we simulate an XZZX memory experiment using the effective error channel of the gadget  estimated via QPT, and we compare it with the standard transpilation to a CZ gate that has a higher bias degradation.

\begin{figure}
    \centering
    \includegraphics[width=\linewidth]{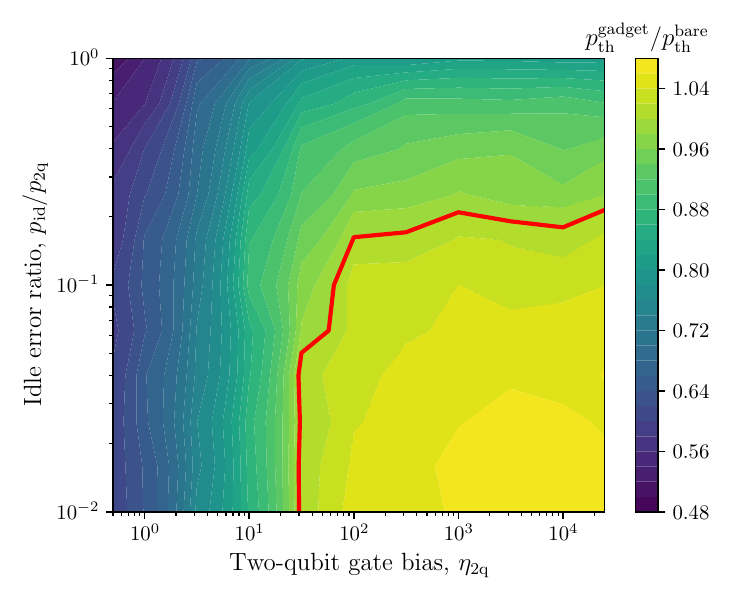}
    \caption{{\bf XZZX threshold ratio for encoded CNOT}: comparison of the XZZX code threshold using encoded or bare CNOT gate for syndrome extraction, as a function of the two-qubit gate bias $\eta_\text{2q}$ and the ratio of dephasing errors to gate errors (where $p\equiv p_\text{2q}=p_\text{1q}$). The encoded CNOT outperforms the traditional syndrome extraction circuit for the yellow region delimited by the red curve.
    }
    \label{fig:contour_gadget}
\end{figure}

In Fig.~\ref{fig:contour_gadget}, we show the ratio of the QEC  thresholds of the XZZX  quantum memory obtained by simulating noisy syndrome-extraction circuits that use the bias-filtering gadget with those that use  bare CNOTs. We use a noise model with $p_\text{1q}=p_\text{2q}$ and $\eta_\text{id}=\infty$, which assumes that the idle errors of qubits that are only affected by environmental noise is a pure dephasing channel. We note that this is a very good approximation for trapped-ion physical qubits encoded in different sub-levels of the ground state manifold, as amplitude decay through spontaneous emission of photons is forbidden by selection rules.   Our numerical simulations show an advantage for the bias-filtering CNOT gadget when $\eta_\text{2q}>50$ and $p_\text{id}<p_\text{2q}/5$. If idle errors are sufficiently low (100 times smaller), it is possible to achieve an advantage with slightly noisier two-qubit gates if $p_\text{2q}<1.3p_\text{1q}$. Thus, to secure an advantage with the bias-filtering CNOT gadget, sufficiently low and pure-dephasing idle errors are required, which is again a reasonable assumption for trapped-ion architectures with ground-state qubits and dynamical decoupling pulses that increase the coherence time above several seconds~\cite{wang2021single, nunnerich2025fast}. 
In this regime, it is advantageous to use XZZX codes with the bias-filtering gadget, although the logical advantage saturates at a generally small value. An open question is to test if this filtering procedure becomes  more advantageous for other bias-tailored codes such as LDPC codes, or to devise other more efficient variants of the same filtering idea. 

\begin{figure}
    \centering
    \includegraphics[width=\linewidth]{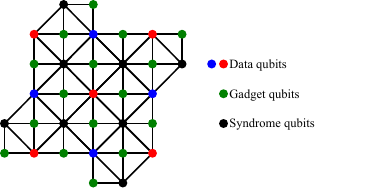}
    \caption{{\bf Connectivity map XZZX+CNOT gadget}: qubit layout and connectivity required to implement a $d=3$ XZZX code with a bias-filtering CNOT gadget for syndrome extraction. Red and blue qubits correspond to data qubits for the XZZX code, and black qubits are ancillas used for syndrome extraction. Green qubits are used to encode each data qubit in a $d=3$ repetition code during syndrome extraction, to implement the CNOT gadget. Edges connect pairs of qubits for which CZ gates are applied.}
    \label{fig:layout-d3}
\end{figure}

Let us close this section by commenting on the connectivity requirements of the proposed bias-filtering CNOT gadget, which can actually be implemented in a experimental architecture such as the one indicated in Fig.~\ref{fig:layout-d3}. This layout requires each qubit to be coupled to 4 or 8 neighbouring qubits. Each data qubit is coupled to the gadget ancillary qubits to allow its encoding into a repetition code before every CNOT gate. Syndrome qubits are coupled to the data qubits in the support of the stabilizer and its associated gadget qubits.

\section{\bf Conclusions and Outlook}\label{sec:conclusions}

In this work, we have investigated the practical limitations of bias-tailored QEC under realistic circuit-level noise. We first compared the bias-tailored XZZX code with anisotropic surface codes, finding that although anisotropic codes can outperform XZZX under idealized code-capacity noise, this advantage largely disappears once realistic syndrome-extraction circuits are included. Under circuit-level noise, both approaches exhibit remarkably similar logical performance with the XZZX code being slightly preferable, indicating that the preservation of the physical noise asymmetry throughout the fault-tolerant circuit is at least as important as the choice of bias-tailoring strategy itself.

Motivated by this observation, we introduced a bias-filtering CNOT gadget designed to reduce bias degradation during syndrome extraction. Our approach uses a temporary repetition code  only during the execution of the bias-degrading CNOT gates. This allows the gadget to be naturally integrated within local topological-code architectures, while preserving their sparse syndrome structure and compatibility with efficient decoding algorithms. Through logical Pauli channel characterization of this bias-filtering gadget, and explicit circuit-level simulations, we identified a regime of high bias and low idle-error rates in which the gadget improves the threshold of the XZZX code by trading additional bias-preserving faults for a suppression of more detrimental bias-degrading error events.

 Looking forward, it could be interesting to extend these ideas to other fault-tolerant primitives, including lattice surgery, or other codes such as LDPC codes, as well as to investigate alternative strategies of bias-aware circuit design, compilation and decoding. Understanding and controlling the propagation of noise asymmetries through fault-tolerant circuits may prove as important as the development of new bias-tailored codes themselves.

\acknowledgments
We gratefully acknowledge support by the European Union’s Horizon Europe research and innovation program under Grant Agreement Number
101114305 (“MILLENION-SGA1” EU Project), by the Office of the
Director of National Intelligence (ODNI), Intelligence Advanced Research Projects Activity (IARPA), under the Entangled Logical Qubits program through Cooperative Agreement Number W911NF-23-2-0216. We are  also supported by PID2021-127726NB-I00 and
PID2024-161474NB-I00 (MCIU/AEI/FEDER,UE) from
QUITEMAD-CM TEC-2024/COM-84, from the Grant
IFT Centro de Excelencia Severo Ochoa CEX2020-001007-
S funded by MCIN/AEI/10.13039/501100011033, and
from the CSIC Research Platform on Quantum Technologies PTI-001. C.B. acknowledges support from Spanish Ministry of Science, Innovation and Universities under grant FPU24/01105.

\bibliography{biblio}

@article{PhysRevLett.78.390,
  title = {Complete Characterization of a Quantum Process: The Two-Bit Quantum Gate},
  author = {Poyatos, J. F. and Cirac, J. I. and Zoller, P.},
  journal = {Phys. Rev. Lett.},
  volume = {78},
  issue = {2},
  pages = {390--393},
  numpages = {0},
  year = {1997},
  month = {Jan},
  publisher = {American Physical Society},
  doi = {10.1103/PhysRevLett.78.390},
  url = {https://link.aps.org/doi/10.1103/PhysRevLett.78.390}
}

@article{PhysRevA.63.020101,
  title = {Maximum-likelihood estimation of quantum processes},
  author = {Fiur\'a\ifmmode \check{s}\else \v{s}\fi{}ek, Jarom\'{\i}r and Hradil, Zden\ifmmode \check{e}\else \v{e}\fi{}k},
  journal = {Phys. Rev. A},
  volume = {63},
  issue = {2},
  pages = {020101(R)},
  numpages = {4},
  year = {2001},
  month = {Jan},
  publisher = {American Physical Society},
  doi = {10.1103/PhysRevA.63.020101},
  url = {https://link.aps.org/doi/10.1103/PhysRevA.63.020101}
}

@article{Chamberland2018flagfaulttolerant,
  doi = {10.22331/q-2018-02-08-53},
  url = {https://doi.org/10.22331/q-2018-02-08-53},
  title = {Flag fault-tolerant error correction with arbitrary distance codes},
  author = {Chamberland, Christopher and Beverland, Michael E.},
  journal = {{Quantum}},
  issn = {2521-327X},
  publisher = {{Verein zur F{\"{o}}rderung des Open Access Publizierens in den Quantenwissenschaften}},
  volume = {2},
  pages = {53},
  month = feb,
  year = {2018}
}

@article{PhysRevLett.121.050502,
  title = {Quantum Error Correction with Only Two Extra Qubits},
  author = {Chao, Rui and Reichardt, Ben W.},
  journal = {Phys. Rev. Lett.},
  volume = {121},
  issue = {5},
  pages = {050502},
  numpages = {5},
  year = {2018},
  month = {Aug},
  publisher = {American Physical Society},
  doi = {10.1103/PhysRevLett.121.050502},
  url = {https://link.aps.org/doi/10.1103/PhysRevLett.121.050502}
}

@Article{Chao2018,
author={Chao, Rui
and Reichardt, Ben W.},
title={Fault-tolerant quantum computation with few qubits},
journal={npj Quantum Information},
year={2018},
month={Sep},
day={12},
volume={4},
number={1},
pages={42},
issn={2056-6387},
doi={10.1038/s41534-018-0085-z},
url={https://doi.org/10.1038/s41534-018-0085-z}
}

@article{PhysRevLett.109.160503,
  title = {High Threshold Error Correction for the Surface Code},
  author = {Wootton, James R. and Loss, Daniel},
  journal = {Phys. Rev. Lett.},
  volume = {109},
  issue = {16},
  pages = {160503},
  numpages = {5},
  year = {2012},
  month = {Oct},
  publisher = {American Physical Society},
  doi = {10.1103/PhysRevLett.109.160503},
  url = {https://link.aps.org/doi/10.1103/PhysRevLett.109.160503}
}

@article{PhysRevX.2.021004,
  title = {Strong Resilience of Topological Codes to Depolarization},
  author = {Bombin, H. and Andrist, Ruben S. and Ohzeki, Masayuki and Katzgraber, Helmut G. and Martin-Delgado, M. A.},
  journal = {Phys. Rev. X},
  volume = {2},
  issue = {2},
  pages = {021004},
  numpages = {10},
  year = {2012},
  month = {Apr},
  publisher = {American Physical Society},
  doi = {10.1103/PhysRevX.2.021004},
  url = {https://link.aps.org/doi/10.1103/PhysRevX.2.021004}
}

@article{Benito2025comparativestudyof,
  doi = {10.22331/q-2025-02-06-1623},
  url = {https://doi.org/10.22331/q-2025-02-06-1623},
  title = {Comparative study of quantum error correction strategies for the heavy-hexagonal lattice},
  author = {Benito, C{\'{e}}sar and L{\'{o}}pez, Esperanza and Peropadre, Borja and Bermudez, Alejandro},
  journal = {{Quantum}},
  issn = {2521-327X},
  publisher = {{Verein zur F{\"{o}}rderung des Open Access Publizierens in den Quantenwissenschaften}},
  volume = {9},
  pages = {1623},
  month = feb,
  year = {2025}
}

@article{PhysRevA.99.022330,
  title = {Transversality and lattice surgery: Exploring realistic routes toward coupled logical qubits with trapped-ion quantum processors},
  author = {Guti\'errez, M. and M\"uller, M. and Berm\'udez, A.},
  journal = {Phys. Rev. A},
  volume = {99},
  issue = {2},
  pages = {022330},
  numpages = {29},
  year = {2019},
  month = {Feb},
  publisher = {American Physical Society},
  doi = {10.1103/PhysRevA.99.022330},
  url = {https://link.aps.org/doi/10.1103/PhysRevA.99.022330}
}

@article{PhysRevA.100.062307,
  title = {Fault-tolerant protection of near-term trapped-ion topological qubits under realistic noise sources},
  author = {Bermudez, A. and Xu, X. and Guti\'errez, M. and Benjamin, S. C. and M\"uller, M.},
  journal = {Phys. Rev. A},
  volume = {100},
  issue = {6},
  pages = {062307},
  numpages = {23},
  year = {2019},
  month = {Dec},
  publisher = {American Physical Society},
  doi = {10.1103/PhysRevA.100.062307},
  url = {https://link.aps.org/doi/10.1103/PhysRevA.100.062307}
}

@article{PhysRevA.90.062320,
  title = {Low-distance surface codes under realistic quantum noise},
  author = {Tomita, Yu and Svore, Krysta M.},
  journal = {Phys. Rev. A},
  volume = {90},
  issue = {6},
  pages = {062320},
  numpages = {15},
  year = {2014},
  month = {Dec},
  publisher = {American Physical Society},
  doi = {10.1103/PhysRevA.90.062320},
  url = {https://link.aps.org/doi/10.1103/PhysRevA.90.062320}
}

@misc{benito2025scalingroadmapmodulartrappedion,
      title={Scaling roadmap for modular trapped-ion QEC and lattice-surgery teleportation}, 
      author={César Benito and Alfredo Ricci Vasquez and Jonathan Home and Karan K. Mehta and Thomas Monz and Markus Müller and Alejandro Bermudez},
      year={2025},
      eprint={2512.20435},
      archivePrefix={arXiv},
      primaryClass={quant-ph},
      url={https://arxiv.org/abs/2512.20435}, 
}

@article{PhysRevX.7.041061,
  title = {Assessing the Progress of Trapped-Ion Processors Towards Fault-Tolerant Quantum Computation},
  author = {Bermudez, A. and Xu, X. and Nigmatullin, R. and O'Gorman, J. and Negnevitsky, V. and Schindler, P. and Monz, T. and Poschinger, U. G. and Hempel, C. and Home, J. and Schmidt-Kaler, F. and Biercuk, M. and Blatt, R. and Benjamin, S. and M\"uller, M.},
  journal = {Phys. Rev. X},
  volume = {7},
  issue = {4},
  pages = {041061},
  numpages = {41},
  year = {2017},
  month = {Dec},
  publisher = {American Physical Society},
  doi = {10.1103/PhysRevX.7.041061},
  url = {https://link.aps.org/doi/10.1103/PhysRevX.7.041061}
}

@Article{Kielpinski2002,
author={Kielpinski, D.
and Monroe, C.
and Wineland, D. J.},
title={Architecture for a large-scale ion-trap quantum computer},
journal={Nature},
year={2002},
month={Jun},
day={01},
volume={417},
number={6890},
pages={709-711},
issn={1476-4687},
doi={10.1038/nature00784},
url={https://doi.org/10.1038/nature00784}
}

@Article{Pino2021,
author={Pino, J. M.
and Dreiling, J. M.
and Figgatt, C.
and Gaebler, J. P.
and Moses, S. A.
and Allman, M. S.
and Baldwin, C. H.
and Foss-Feig, M.
and Hayes, D.
and Mayer, K.
and Ryan-Anderson, C.
and Neyenhuis, B.},
title={Demonstration of the trapped-ion quantum CCD computer architecture},
journal={Nature},
year={2021},
month={Apr},
day={01},
volume={592},
number={7853},
pages={209-213},
issn={1476-4687},
doi={10.1038/s41586-021-03318-4},
url={https://doi.org/10.1038/s41586-021-03318-4}
}

@Article{Lacroix2025,
author={Lacroix, N.
and Bourassa, A.
and Heras, F. J. H.
and Zhang, L. M.
and Bausch, J.
and Senior, A. W.
and Edlich, T.
and Shutty, N.
and Sivak, V.
and Bengtsson, A.
and McEwen, M.
and Higgott, O.
and Kafri, D.
and Claes, J.
and Morvan, A.
and Chen, Z.
and Zalcman, A.
and Madhuk, S.
and others},
title={Scaling and logic in the colour code on a superconducting quantum processor},
journal={Nature},
year={2025},
month={Sep},
day={01},
volume={645},
number={8081},
pages={614-619},
issn={1476-4687},
doi={10.1038/s41586-025-09061-4},
url={https://doi.org/10.1038/s41586-025-09061-4}
}

@article{
doi:10.1126/science.adp6016,
author = {C. Ryan-Anderson  and N. C. Brown  and C. H. Baldwin  and J. M. Dreiling  and C. Foltz  and J. P. Gaebler  and T. M. Gatterman  and N. Hewitt  and C. Holliman  and C. V. Horst  and J. Johansen  and D. Lucchetti  and T. Mengle  and M. Matheny  and Y. Matsuoka  and K. Mayer  and M. Mills  and S. A. Moses  and B. Neyenhuis  and J. Pino  and P. Siegfried  and R. P. Stutz  and J. Walker  and D. Hayes },
title = {High-fidelity teleportation of a logical qubit using transversal gates and lattice surgery},
journal = {Science},
volume = {385},
number = {6715},
pages = {1327-1331},
year = {2024},
doi = {10.1126/science.adp6016},
URL = {https://www.science.org/doi/abs/10.1126/science.adp6016}}

@Article{SalesRodriguez2025,
author={Sales Rodriguez, Pedro
and Robinson, John M.
and Jepsen, Paul Niklas
and He, Zhiyang
and Duckering, Casey
and Zhao, Chen
and Wu, Kai-Hsin
and Campo, Joseph
and Bagnall, Kevin
and Kwon, Minho
and Karolyshyn, Thomas
and Weinberg, Phillip
and Cain, Madelyn
and Evered, Simon J.
and Geim, Alexandra A.
and others},
title={Experimental demonstration of logical magic state distillation},
journal={Nature},
year={2025},
month={Sep},
day={01},
volume={645},
number={8081},
pages={620-625},
issn={1476-4687},
doi={10.1038/s41586-025-09367-3},
url={https://doi.org/10.1038/s41586-025-09367-3}
}

@misc{perlin2026faulttolerantexecutionerrorcorrectedquantum,
      title={Fault-tolerant execution of error-corrected quantum algorithms}, 
      author={Michael A. Perlin and Zichang He and Anthony Alexiades Armenakas and Pablo Andres-Martinez and Tianyi Hao and Dylan Herman and Yuwei Jin and Karl Mayer and Chris Self and David Amaro and Ciaran Ryan-Anderson and Ruslan Shaydulin},
      year={2026},
      eprint={2603.04584},
      archivePrefix={arXiv},
      primaryClass={quant-ph},
      url={https://arxiv.org/abs/2603.04584}, 
}

@misc{mayer2024benchmarkinglogicalthreequbitquantum,
      title={Benchmarking logical three-qubit quantum Fourier transform encoded in the Steane code on a trapped-ion quantum computer}, 
      author={Karl Mayer and Ciarán Ryan-Anderson and Natalie Brown and Elijah Durso-Sabina and Charles H. Baldwin and David Hayes and Joan M. Dreiling and Cameron Foltz and John P. Gaebler and Thomas M. Gatterman and Justin A. Gerber and Kevin Gilmore and Dan Gresh and Nathan Hewitt and Chandler V. Horst and Jacob Johansen and Tanner Mengle and Michael Mills and Steven A. Moses and Peter E. Siegfried and Brian Neyenhuis and Juan Pino and Russell Stutz},
      year={2024},
      eprint={2404.08616},
      archivePrefix={arXiv},
      primaryClass={quant-ph},
      url={https://arxiv.org/abs/2404.08616}, 
}

@article{PhysRevA.63.054104,
  title = {Maximum-likelihood reconstruction of completely positive maps},
  author = {Sacchi, Massimiliano F.},
  journal = {Phys. Rev. A},
  volume = {63},
  issue = {5},
  pages = {054104},
  numpages = {4},
  year = {2001},
  month = {Apr},
  publisher = {American Physical Society},
  doi = {10.1103/PhysRevA.63.054104},
  url = {https://link.aps.org/doi/10.1103/PhysRevA.63.054104}
}

@article{PhysRevA.68.012305,
  title = {Quantum inference of states and processes},
  author = {Je\ifmmode \check{z}\else \v{z}\fi{}ek, Miroslav and Fiur\'a\ifmmode \check{s}\else \v{s}\fi{}ek, Jarom\'{\i}r and Hradil, Zden\ifmmode \check{e}\else \v{e}\fi{}k},
  journal = {Phys. Rev. A},
  volume = {68},
  issue = {1},
  pages = {012305},
  numpages = {7},
  year = {2003},
  month = {Jul},
  publisher = {American Physical Society},
  doi = {10.1103/PhysRevA.68.012305},
  url = {https://link.aps.org/doi/10.1103/PhysRevA.68.012305}
}

@article{Chuang01111997,
author = {Isaac L. Chuang and M. A. Nielsen},
title = {Prescription for experimental determination of the dynamics of a quantum black box},
journal = {Journal of Modern Optics},
volume = {44},
number = {11-12},
pages = {2455--2467},
year = {1997},
publisher = {Taylor \& Francis},
doi = {10.1080/09500349708231894},


URL = { 
    
    
        https://www.tandfonline.com/doi/abs/10.1080/09500349708231894
    

}

}

@article{PhysRevA.70.052328,
  title = {Improved simulation of stabilizer circuits},
  author = {Aaronson, Scott and Gottesman, Daniel},
  journal = {Phys. Rev. A},
  volume = {70},
  issue = {5},
  pages = {052328},
  numpages = {14},
  year = {2004},
  month = {Nov},
  publisher = {American Physical Society},
  doi = {10.1103/PhysRevA.70.052328},
  url = {https://link.aps.org/doi/10.1103/PhysRevA.70.052328}
}

@book{Nielsen_Chuang_2010, place={Cambridge}, title={Quantum Computation and Quantum Information: 10th Anniversary Edition}, publisher={Cambridge University Press}, author={Nielsen, Michael A. and Chuang, Isaac L.}, year={2010}, doi={10.1017/CBO9780511976667}}

@misc{martinez2025leveragingbiasednoiseefficient,
      title={Leveraging biased noise for more efficient quantum error correction at the circuit-level with two-level qubits}, 
      author={Josu Etxezarreta Martinez and Paul Schnabl and Javier Oliva del Moral and Reza Dastbasteh and Pedro M. Crespo and Ruben M. Otxoa},
      year={2025},
      eprint={2505.17718},
      archivePrefix={arXiv},
      primaryClass={quant-ph},
      url={https://arxiv.org/abs/2505.17718}, 
}

@misc{aharonov1999faulttolerantquantumcomputationconstant,
      title={Fault-Tolerant Quantum Computation With Constant Error Rate}, 
      author={Dorit Aharonov and Michael Ben-Or},
      year={1999},
      eprint={quant-ph/9906129},
      archivePrefix={arXiv},
      primaryClass={quant-ph},
      url={https://arxiv.org/abs/quant-ph/9906129}, 
}

@article{6qbg-xslr,
  title = {Quantum $XYZ$ cyclic codes for biased noise},
  author = {Liang, Zhipeng and Yang, Fusheng and Yi, Zhengzhong and Wang, Xuan},
  journal = {Phys. Rev. A},
  volume = {112},
  issue = {1},
  pages = {012402},
  numpages = {14},
  year = {2025},
  month = {Jul},
  publisher = {American Physical Society},
  doi = {10.1103/6qbg-xslr},
  url = {https://link.aps.org/doi/10.1103/6qbg-xslr}
}

@article{rrfk-1b51,
  title = {High-dimensional quantum $XYZ$ product codes for biased noise},
  author = {Liang, Zhipeng and Yi, Zhengzhong and Yang, Fusheng and Chen, Jiahan and Wang, Zicheng and Wang, Xuan},
  journal = {Phys. Rev. A},
  volume = {112},
  issue = {5},
  pages = {052439},
  numpages = {19},
  year = {2025},
  month = {Nov},
  publisher = {American Physical Society},
  doi = {10.1103/rrfk-1b51},
  url = {https://link.aps.org/doi/10.1103/rrfk-1b51}
}

@Article{Setiawan2025,
author={Setiawan, F.
and McLauchlan, Campbell},
title={Tailoring dynamical codes for biased noise: the X3Z3 Floquet code},
journal={npj Quantum Information},
year={2025},
month={Sep},
day={22},
volume={11},
number={1},
pages={149},
issn={2056-6387},
doi={10.1038/s41534-025-01074-1},
url={https://doi.org/10.1038/s41534-025-01074-1}
}

@article{Wang:2009bfa,
    author = "Wang, David S. and Fowler, Austin G. and Stephens, Ashley M. and Hollenberg, Lloyd C. L.",
    title = "{Threshold error rates for the toric and planar codes}",
    eprint = "0905.0531",
    archivePrefix = "arXiv",
    primaryClass = "quant-ph",
    doi = "10.26421/QIC10.5-6-6",
    journal = "Quant. Inf. Comput.",
    volume = "10",
    number = "5-6",
    pages = "0456--0469",
    year = "2010"
}

@article{10.1098/rspa.1998.0166,
    author = {Knill, Emanuel and Laflamme, Raymond and Zurek, Wojciech H.},
    title = {Resilient quantum computation: error models and thresholds},
    journal = {Proceedings of the Royal Society A: Mathematical, Physical and Engineering Sciences},
    volume = {454},
    number = {1969},
    pages = {365-384},
    year = {1998},
    month = {01},
    issn = {1364-5021},
    doi = {10.1098/rspa.1998.0166},
    url = {https://doi.org/10.1098/rspa.1998.0166},
    eprint = {https://royalsocietypublishing.org/rspa/article-pdf/454/1969/365/633972/rspa.1998.0166.pdf},
}

@misc{landahl2011faulttolerantquantumcomputingcolor,
      title={Fault-tolerant quantum computing with color codes}, 
      author={Andrew J. Landahl and Jonas T. Anderson and Patrick R. Rice},
      year={2011},
      eprint={1108.5738},
      archivePrefix={arXiv},
      primaryClass={quant-ph},
      url={https://arxiv.org/abs/1108.5738}, 
}

@article{PhysRevA.86.032324,
  title = {Surface codes: Towards practical large-scale quantum computation},
  author = {Fowler, Austin G. and Mariantoni, Matteo and Martinis, John M. and Cleland, Andrew N.},
  journal = {Phys. Rev. A},
  volume = {86},
  issue = {3},
  pages = {032324},
  numpages = {48},
  year = {2012},
  month = {Sep},
  publisher = {American Physical Society},
  doi = {10.1103/PhysRevA.86.032324},
  url = {https://link.aps.org/doi/10.1103/PhysRevA.86.032324}
}

@article{RevModPhys.87.307,
  title = {Quantum error correction for quantum memories},
  author = {Terhal, Barbara M.},
  journal = {Rev. Mod. Phys.},
  volume = {87},
  issue = {2},
  pages = {307--346},
  numpages = {40},
  year = {2015},
  month = {Apr},
  publisher = {American Physical Society},
  doi = {10.1103/RevModPhys.87.307},
  url = {https://link.aps.org/doi/10.1103/RevModPhys.87.307}
}

@misc{Eisert:2025ytq,
    author = {Eisert, Jens and Preskill, John},
    title = {Mind the gaps: The fraught road to quantum advantage},
    eprint = {2510.19928},
    archivePrefix = {arXiv},
    primaryClass = {quant-ph},
    month = {Oct},
    year = {2025}
}

@book{Dalzell_McArdle, place={Cambridge}, title={Quantum Algorithms: A Survey of Applications and End-to-end Complexities}, publisher={Cambridge University Press}, author={Dalzell, Alexander M. and McArdle, Sam and Berta, Mario and Bienias, Przemyslaw and Chen, Chi-Fang and Gilyén, András and Hann, Connor T. and Kastoryano, Michael J. and Khabiboulline, Emil T. and Kubica, Aleksander and et al.}, year={2025}, doi={10.1017/9781009639651}}

@article{lee2021rectangular,
  title={Rectangular surface code under biased noise},
  author={Lee, Jonghyun and Park, Jooyoun and Heo, Jun},
  journal={Quantum Information Processing},
  volume={20},
  number={7},
  pages={231},
  year={2021},
  publisher={Springer},
  url={https://link.springer.com/article/10.1007/s11128-021-03130-z}
}

@INPROCEEDINGS{Shor1996,
  author={Shor, P.W.},
  booktitle={Proceedings of 37th Conference on Foundations of Computer Science}, 
  title={Fault-tolerant quantum computation}, 
  year={1996},
  volume={},
  number={},
  pages={56-65},
  doi={10.1109/SFCS.1996.548464}}

@Article{Google2025below,
author={AI, Google Quantum
and {Collaborators}},
title={Quantum error correction below the surface code threshold},
journal={Nature},
year={2025},
month={Feb},
day={01},
volume={638},
number={8052},
pages={920-926},
issn={1476-4687},
doi={10.1038/s41586-024-08449-y},
url={https://doi.org/10.1038/s41586-024-08449-y}
}

@article{Postler_2022,
   title={Demonstration of fault-tolerant universal quantum gate operations},
   volume={605},
   ISSN={1476-4687},
   url={http://dx.doi.org/10.1038/s41586-022-04721-1},
   DOI={10.1038/s41586-022-04721-1},
   number={7911},
   journal={Nature},
   publisher={Springer Science and Business Media LLC},
   author={Postler, Lukas and Heu{\ss}en, Sascha and Pogorelov, Ivan and Rispler, Manuel and Feldker, Thomas and Meth, Michael and Marciniak, Christian D. and Stricker, Roman and Ringbauer, Martin and Blatt, Rainer and Schindler, Philipp and Müller, Markus and Monz, Thomas},
   year={2022},
   month=May, pages={675–680} }

@Article{Bluvstein2026,
author={Bluvstein, Dolev
and Geim, Alexandra A.
and Li, Sophie H.
and Evered, Simon J.
and Bonilla Ataides, J. Pablo
and Baranes, Gefen
and Gu, Andi
and Manovitz, Tom
and Xu, Muqing
and Kalinowski, Marcin
and Majidy, Shayan
and Kokail, Christian
and Maskara, Nishad
and Trapp, Elias C.
and Stewart, Luke M.
and Hollerith, Simon
and Zhou, Hengyun
and Gullans, Michael J.
and Yelin, Susanne F.
and Greiner, Markus
and Vuleti{\'{c}}, Vladan
and Cain, Madelyn
and Lukin, Mikhail D.},
title={A fault-tolerant neutral-atom architecture for universal quantum computation},
journal={Nature},
year={2026},
month={Jan},
day={01},
volume={649},
number={8095},
pages={39-46},
issn={1476-4687},
doi={10.1038/s41586-025-09848-5},
url={https://doi.org/10.1038/s41586-025-09848-5}
}

@article{PhysRevA.52.R2493,
  title = {Scheme for reducing decoherence in quantum computer memory},
  author = {Shor, Peter W.},
  journal = {Phys. Rev. A},
  volume = {52},
  issue = {4},
  pages = {R2493--R2496},
  numpages = {0},
  year = {1995},
  month = {Oct},
  publisher = {American Physical Society},
  doi = {10.1103/PhysRevA.52.R2493},
  url = {https://link.aps.org/doi/10.1103/PhysRevA.52.R2493}
}

@article{10.1098/rspa.1996.0136,
    author = {Steane, Andrew},
    title = {Multiple-particle interference and quantum error correction},
    journal = {Proceedings of the Royal Society A: Mathematical, Physical and Engineering Sciences},
    volume = {452},
    number = {1954},
    pages = {2551-2577},
    year = {1996},
    month = {11},
    issn = {1364-5021},
    doi = {10.1098/rspa.1996.0136},
    url = {https://doi.org/10.1098/rspa.1996.0136},
    eprint = {https://royalsocietypublishing.org/rspa/article-pdf/452/1954/2551/998878/rspa.1996.0136.pdf},
}

@ARTICLE{6671468,
  author={Tillich, Jean-Pierre and Zémor, Gilles},
  journal={IEEE Transactions on Information Theory}, 
  title={Quantum LDPC Codes With Positive Rate and Minimum Distance Proportional to the Square Root of the Blocklength}, 
  year={2014},
  volume={60},
  number={2},
  pages={1193-1202},
  doi={10.1109/TIT.2013.2292061}}

@article{PhysRevLett.84.2525,
  title = {Theory of Quantum Error Correction for General Noise},
  author = {Knill, Emanuel and Laflamme, Raymond and Viola, Lorenza},
  journal = {Phys. Rev. Lett.},
  volume = {84},
  issue = {11},
  pages = {2525--2528},
  numpages = {0},
  year = {2000},
  month = {Mar},
  publisher = {American Physical Society},
  doi = {10.1103/PhysRevLett.84.2525},
  url = {https://link.aps.org/doi/10.1103/PhysRevLett.84.2525}
}

@article{PhysRevA.57.127,
  title = {Theory of fault-tolerant quantum computation},
  author = {Gottesman, Daniel},
  journal = {Phys. Rev. A},
  volume = {57},
  issue = {1},
  pages = {127--137},
  numpages = {0},
  year = {1998},
  month = {Jan},
  publisher = {American Physical Society},
  doi = {10.1103/PhysRevA.57.127},
  url = {https://link.aps.org/doi/10.1103/PhysRevA.57.127}
}

@article{KITAEV20032,
title = {Fault-tolerant quantum computation by anyons},
journal = {Annals of Physics},
volume = {303},
number = {1},
pages = {2-30},
year = {2003},
issn = {0003-4916},
doi = {https://doi.org/10.1016/S0003-4916(02)00018-0},
url = {https://www.sciencedirect.com/science/article/pii/S0003491602000180},
author = {A.Yu. Kitaev}
}

@article{10.1063/1.1499754,
    author = {Dennis, Eric and Kitaev, Alexei and Landahl, Andrew and Preskill, John},
    title = {Topological quantum memory},
    journal = {Journal of Mathematical Physics},
    volume = {43},
    number = {9},
    pages = {4452-4505},
    year = {2002},
    month = {09},
    issn = {0022-2488},
    doi = {10.1063/1.1499754},
    url = {https://doi.org/10.1063/1.1499754}
}

@misc{bravyi1998quantumcodeslatticeboundary,
      title={Quantum codes on a lattice with boundary}, 
      author={S. B. Bravyi and A. Yu. Kitaev},
      year={1998},
      eprint={quant-ph/9811052},
      archivePrefix={arXiv},
      primaryClass={quant-ph},
      url={https://arxiv.org/abs/quant-ph/9811052}, 
}

@article{PhysRevLett.97.180501,
  title = {Topological Quantum Distillation},
  author = {Bombin, H. and Martin-Delgado, M. A.},
  journal = {Phys. Rev. Lett.},
  volume = {97},
  issue = {18},
  pages = {180501},
  numpages = {4},
  year = {2006},
  month = {Oct},
  publisher = {American Physical Society},
  doi = {10.1103/PhysRevLett.97.180501},
  url = {https://link.aps.org/doi/10.1103/PhysRevLett.97.180501}
}

@Article{BonillaAtaides2021,
author={Bonilla Ataides, J. Pablo
and Tuckett, David K.
and Bartlett, Stephen D.
and Flammia, Steven T.
and Brown, Benjamin J.},
title={The XZZX surface code},
journal={Nature Communications},
year={2021},
month={Apr},
day={12},
volume={12},
number={1},
pages={2172},
issn={2041-1723},
doi={10.1038/s41467-021-22274-1},
url={https://doi.org/10.1038/s41467-021-22274-1}
}

@article{PhysRevA.76.012305,
  title = {Optimal resources for topological two-dimensional stabilizer codes: Comparative study},
  author = {Bombin, H. and Martin-Delgado, M. A.},
  journal = {Phys. Rev. A},
  volume = {76},
  issue = {1},
  pages = {012305},
  numpages = {6},
  year = {2007},
  month = {Jul},
  publisher = {American Physical Society},
  doi = {10.1103/PhysRevA.76.012305},
  url = {https://link.aps.org/doi/10.1103/PhysRevA.76.012305}
}

@article{PhysRevA.78.052331,
  title = {Fault-tolerant quantum computation against biased noise},
  author = {Aliferis, Panos and Preskill, John},
  journal = {Phys. Rev. A},
  volume = {78},
  issue = {5},
  pages = {052331},
  numpages = {9},
  year = {2008},
  month = {Nov},
  publisher = {American Physical Society},
  doi = {10.1103/PhysRevA.78.052331},
  url = {https://link.aps.org/doi/10.1103/PhysRevA.78.052331}
}

@article{PhysRevLett.123.110503,
  title = {Probing Qubit Memory Errors at the Part-per-Million Level},
  author = {Sepiol, M. A. and Hughes, A. C. and Tarlton, J. E. and Nadlinger, D. P. and Ballance, T. G. and Ballance, C. J. and Harty, T. P. and Steane, A. M. and Goodwin, J. F. and Lucas, D. M.},
  journal = {Phys. Rev. Lett.},
  volume = {123},
  issue = {11},
  pages = {110503},
  numpages = {5},
  year = {2019},
  month = {Sep},
  publisher = {American Physical Society},
  doi = {10.1103/PhysRevLett.123.110503},
  url = {https://link.aps.org/doi/10.1103/PhysRevLett.123.110503}
}

@article{Roffe2023biastailoredquantum,
  doi = {10.22331/q-2023-05-15-1005},
  url = {https://doi.org/10.22331/q-2023-05-15-1005},
  title = {Bias-tailored quantum {LDPC} codes},
  author = {Roffe, Joschka and Cohen, Lawrence Z. and Quintavalle, Armanda O. and Chandra, Daryus and Campbell, Earl T.},
  journal = {{Quantum}},
  issn = {2521-327X},
  publisher = {{Verein zur F{\"{o}}rderung des Open Access Publizierens in den Quantenwissenschaften}},
  volume = {7},
  pages = {1005},
  month = may,
  year = {2023}
}

@article{PhysRevResearch.5.013035,
  title = {Tailored XZZX codes for biased noise},
  author = {Xu, Qian and Mannucci, Nam and Seif, Alireza and Kubica, Aleksander and Flammia, Steven T. and Jiang, Liang},
  journal = {Phys. Rev. Res.},
  volume = {5},
  issue = {1},
  pages = {013035},
  numpages = {15},
  year = {2023},
  month = {Jan},
  publisher = {American Physical Society},
  doi = {10.1103/PhysRevResearch.5.013035},
  url = {https://link.aps.org/doi/10.1103/PhysRevResearch.5.013035}
}

@article{PhysRevX.9.041031,
  title = {Tailoring Surface Codes for Highly Biased Noise},
  author = {Tuckett, David K. and Darmawan, Andrew S. and Chubb, Christopher T. and Bravyi, Sergey and Bartlett, Stephen D. and Flammia, Steven T.},
  journal = {Phys. Rev. X},
  volume = {9},
  issue = {4},
  pages = {041031},
  numpages = {22},
  year = {2019},
  month = {Nov},
  publisher = {American Physical Society},
  doi = {10.1103/PhysRevX.9.041031},
  url = {https://link.aps.org/doi/10.1103/PhysRevX.9.041031}
}

@misc{wu2025biastailoredsingleshotquantumldpc,
      title={Bias-tailored single-shot quantum LDPC codes}, 
      author={Shixin Wu and Todd A. Brun and Daniel A. Lidar},
      year={2025},
      eprint={2507.02239},
      archivePrefix={arXiv},
      primaryClass={quant-ph},
      url={https://arxiv.org/abs/2507.02239}, 
}

@article{Gidney2021stimfaststabilizer,
  doi = {10.22331/q-2021-07-06-497},
  url = {https://doi.org/10.22331/q-2021-07-06-497},
  title = {Stim: a fast stabilizer circuit simulator},
  author = {Gidney, Craig},
  journal = {{Quantum}},
  issn = {2521-327X},
  publisher = {{Verein zur F{\"{o}}rderung des Open Access Publizierens in den Quantenwissenschaften}},
  volume = {5},
  pages = {497},
  month = jul,
  year = {2021}
}

@article{PhysRevX.9.041053,
  title = {Repetition Cat Qubits for Fault-Tolerant Quantum Computation},
  author = {Guillaud, J\'er\'emie and Mirrahimi, Mazyar},
  journal = {Phys. Rev. X},
  volume = {9},
  issue = {4},
  pages = {041053},
  numpages = {23},
  year = {2019},
  month = {Dec},
  publisher = {American Physical Society},
  doi = {10.1103/PhysRevX.9.041053},
  url = {https://link.aps.org/doi/10.1103/PhysRevX.9.041053}
}

@article{PhysRevA.92.062309,
  title = {Reducing the overhead for quantum computation when noise is biased},
  author = {Webster, Paul and Bartlett, Stephen D. and Poulin, David},
  journal = {Phys. Rev. A},
  volume = {92},
  issue = {6},
  pages = {062309},
  numpages = {8},
  year = {2015},
  month = {Dec},
  publisher = {American Physical Society},
  doi = {10.1103/PhysRevA.92.062309},
  url = {https://link.aps.org/doi/10.1103/PhysRevA.92.062309}
}

@article{lescanne2020exponential,
  title={Exponential suppression of bit-flips in a qubit encoded in an oscillator},
  author={Lescanne, Rapha{\"e}l and Villiers, Marius and Peronnin, Th{\'e}au and Sarlette, Alain and Delbecq, Matthieu and Huard, Benjamin and Kontos, Takis and Mirrahimi, Mazyar and Leghtas, Zaki},
  journal={Nature Physics},
  volume={16},
  number={5},
  pages={509--513},
  year={2020},
  publisher={Nature Publishing Group UK London},
  url={https://www.nature.com/articles/s41567-020-0824-x}
}

@article{berdou2023one,
  title={One hundred second bit-flip time in a two-photon dissipative oscillator},
  author={Berdou, Camille and Murani, Anil and Reglade, Ulysse and Smith, William C and Villiers, Marius and Palomo, Jos{\'e} and Rosticher, Michael and Denis, A and Morfin, Pascal and Delbecq, Matthieu and others},
  journal={PRX Quantum},
  volume={4},
  number={2},
  pages={020350},
  year={2023},
  publisher={APS},
  url={https://journals.aps.org/prxquantum/abstract/10.1103/PRXQuantum.4.020350}
}

@article{ding2025quantum,
  title={Quantum control of an oscillator with a Kerr-cat qubit},
  author={Ding, Andy Z and Brock, Benjamin L and Eickbusch, Alec and Koottandavida, Akshay and Frattini, Nicholas E and Corti{\~n}as, Rodrigo G and Joshi, Vidul R and de Graaf, Stijn J and Chapman, Benjamin J and Ganjam, Suhas and others},
  journal={Nature Communications},
  volume={16},
  number={1},
  pages={5279},
  year={2025},
  publisher={Nature Publishing Group UK London},
  url={https://www.nature.com/articles/s41467-025-60352-w}
}

@article{aliferis2008fault,
  title={Fault-tolerant quantum computation against biased noise},
  author={Aliferis, Panos and Preskill, John},
  journal={Physical Review A—Atomic, Molecular, and Optical Physics},
  volume={78},
  number={5},
  pages={052331},
  year={2008},
  publisher={APS},
  url={https://journals.aps.org/pra/abstract/10.1103/PhysRevA.78.052331}
}

@article{seis2023improving,
  title={Improving trapped-ion-qubit memories via code-mediated error-channel balancing},
  author={Seis, Yannick and Brown, Benjamin J and S{\o}rensen, Anders S and Goodwin, Joseph F},
  journal={Physical Review A},
  volume={107},
  number={5},
  pages={052417},
  year={2023},
  publisher={APS},
  url={https://journals.aps.org/pra/abstract/10.1103/PhysRevA.107.052417}
}

@article{takeda2022quantum,
  title={Quantum error correction with silicon spin qubits},
  author={Takeda, Kenta and Noiri, Akito and Nakajima, Takashi and Kobayashi, Takashi and Tarucha, Seigo},
  journal={Nature},
  volume={608},
  number={7924},
  pages={682--686},
  year={2022},
  publisher={Nature Publishing Group UK London},
  url={https://www.nature.com/articles/s41586-022-04986-6}
}

@article{hetenyi2024tailoring,
  title={Tailoring quantum error correction to spin qubits},
  author={Het{\'e}nyi, Bence and Wootton, James R},
  journal={Physical Review A},
  volume={109},
  number={3},
  pages={032433},
  year={2024},
  publisher={APS},
  url={https://journals.aps.org/pra/abstract/10.1103/PhysRevA.109.032433}
}

@article{noiri2022fast,
  title={Fast universal quantum gate above the fault-tolerance threshold in silicon},
  author={Noiri, Akito and Takeda, Kenta and Nakajima, Takashi and Kobayashi, Takashi and Sammak, Amir and Scappucci, Giordano and Tarucha, Seigo},
  journal={Nature},
  volume={601},
  number={7893},
  pages={338--342},
  year={2022},
  publisher={Nature Publishing Group UK London},
  url={https://www.nature.com/articles/s41586-021-04182-y}
}

@article{steinacker2025industry,
  title={Industry-compatible silicon spin-qubit unit cells exceeding 99\% fidelity},
  author={Steinacker, Paul and Dumoulin Stuyck, Nard and Lim, Wee Han and Tanttu, Tuomo and Feng, MengKe and Serrano, Santiago and Nickl, Andreas and Candido, Marco and Cifuentes, Jesus D and Vahapoglu, Ensar and others},
  journal={Nature},
  volume={646},
  number={8083},
  pages={81--87},
  year={2025},
  publisher={Nature Publishing Group UK London},
  url={https://www.nature.com/articles/s41586-025-09531-9}
}

@article{ly5s-tjk2,
  title = {Efficient optical configurations for trapped-ion entangling gates},
  author = {Kolhatkar, Aditya Milind and Mehta, Karan K.},
  journal = {Phys. Rev. A},
  volume = {113},
  issue = {4},
  pages = {042424},
  numpages = {21},
  year = {2026},
  month = {Apr},
  publisher = {American Physical Society},
  doi = {10.1103/ly5s-tjk2},
  url = {https://link.aps.org/doi/10.1103/ly5s-tjk2}
}

@misc{vezvaee2025surfacecodescalingheavyhex,
      title={Surface code scaling on heavy-hex superconducting quantum processors}, 
      author={Arian Vezvaee and Cesar Benito and Mario Morford-Oberst and Alejandro Bermudez and Daniel A. Lidar},
      year={2025},
      eprint={2510.18847},
      archivePrefix={arXiv},
      primaryClass={quant-ph},
      url={https://arxiv.org/abs/2510.18847}, 
}

@article{PhysRevA.54.2614,
  title = {Sending entanglement through noisy quantum channels},
  author = {Schumacher, Benjamin},
  journal = {Phys. Rev. A},
  volume = {54},
  issue = {4},
  pages = {2614--2628},
  numpages = {0},
  year = {1996},
  month = {Oct},
  publisher = {American Physical Society},
  doi = {10.1103/PhysRevA.54.2614},
  url = {https://link.aps.org/doi/10.1103/PhysRevA.54.2614}
}

@article{NIELSEN2002249,
title = {A simple formula for the average gate fidelity of a quantum dynamical operation},
journal = {Physics Letters A},
volume = {303},
number = {4},
pages = {249-252},
year = {2002},
issn = {0375-9601},
doi = {https://doi.org/10.1016/S0375-9601(02)01272-0},
url = {https://www.sciencedirect.com/science/article/pii/S0375960102012720},
author = {Michael A Nielsen}
}

@article{wang2021single,
  title={Single ion qubit with estimated coherence time exceeding one hour},
  author={Wang, Pengfei and Luan, Chun-Yang and Qiao, Mu and Um, Mark and Zhang, Junhua and Wang, Ye and Yuan, Xiao and Gu, Mile and Zhang, Jingning and Kim, Kihwan},
  journal={Nature communications},
  volume={12},
  number={1},
  pages={233},
  year={2021},
  publisher={Nature Publishing Group UK London},
  url={https://www.nature.com/articles/s41467-020-20330-w}
}

@article{nunnerich2025fast,
  title={Fast, robust, and laser-free universal entangling gates for trapped-ion quantum computing},
  author={N{\"u}nnerich, Markus and Cohen, Daniel and Barthel, Patrick and Huber, Patrick H and Niroomand, Dorna and Retzker, Alex and Wunderlich, Christof},
  journal={Physical Review X},
  volume={15},
  number={2},
  pages={021079},
  year={2025},
  publisher={APS},
  url={https://journals.aps.org/prx/abstract/10.1103/PhysRevX.15.021079}
}

@article{Higgott2025sparseblossom,
  doi = {10.22331/q-2025-01-20-1600},
  url = {https://doi.org/10.22331/q-2025-01-20-1600},
  title = {Sparse {B}lossom: correcting a million errors per core second with minimum-weight matching},
  author = {Higgott, Oscar and Gidney, Craig},
  journal = {{Quantum}},
  issn = {2521-327X},
  publisher = {{Verein zur F{\"{o}}rderung des Open Access Publizierens in den Quantenwissenschaften}},
  volume = {9},
  pages = {1600},
  month = jan,
  year = {2025}
}
\appendix
\section{Gadget quantum process tomography (QPT)}\label{app:tomography}
For a two-qubit system, QPT aims at reconstructing the CPTP  channel describing a physically-allowed operation
\begin{equation}
\mathcal E(\rho)
=
\sum_{m,n}\chi_{mn} E_m \rho E_n^\dagger
:\;
\left\{
\begin{array}{c}
\chi \in \mathsf{Pos}(\mathcal H),\\[0.5ex]
\displaystyle\sum_{m,n}\chi_{mn} E_n^\dagger E_m
= \mathbb I_{4}
\end{array}
\right.
\end{equation}
where $\chi$ is the so-called  process matrix, and $\{E_m\}$ is an orthogonal  basis of the space of linear operators $\mathsf{L}(\mathcal{H})$. Choosing the 2-qubit Pauli basis $E_m\in\mathcal{P}_2$, the $\chi$ matrix is a semidefinite positive $16\times16$  matrix subject to the 16 trace constraints, thus being fully specified by $16 (16-1)=240$ independent real parameters.
A typical choice~\cite{Chuang01111997,PhysRevLett.78.390} is to consider the  tensor products of four  Pauli-basis states
    \beq
    \label{eq:IC_initial_set}
    \rho_{0,s}\in\big\{\ket{0}\!\!\bra{0},\ket{1}\!\!\bra{1},\ket{+}\!\!\bra{+},\ket{+{\rm i}}\!\!\bra{+{\rm i}}\big\}^{\!\otimes^2},
    \eeq    
    Additionally, one typically uses local Pauli projectors for the informationally-complete measurements
\beq
\label{eq:local_pauli_proj}
\mu=(\boldsymbol{b},\boldsymbol{m}_{\boldsymbol{b}}),\hspace{2ex}M_{\boldsymbol{b},\boldsymbol{m}_{\boldsymbol{b}}}=\frac{1}{3^2}P_{b_1,m_{b_1}}\otimes P_{b_2,m_{b_2}},
\eeq
where $b_1,b_2\in\{x,y,z\}$ indicate the Pauli basis for each qubit,  and $P_{b_1,m_{b_1}},P_{b_2,m_{b_2}}$ the orthogonal projectors associated to  the  corresponding eigenvalue $m_{b_1},m_{b_2}\in\{ -1,+1\}$. Altogether,  we have $16$ initial states, and $9$ measurement bases with $4$ possible outcomes each, leading to  $4^2\cdot3^2\cdot2^2=576$  probabilities with only $432$ of them being independent. This yields a formal informationally-complete mapping between the process matrix and a probability matrix 
    \beq
    \chi \mapsto \big[q\big]_{\mu s}={\rm Tr}\big\{M_\mu\mathcal{E}(\rho_{0,s})\big\}\approx \sum_{s,\boldsymbol{b},\boldsymbol{m}_{\boldsymbol{b}}} \frac{N_{s,\boldsymbol{b},\boldsymbol{m}_{\boldsymbol{b}}}}{N_{s,  \boldsymbol{b}}}\,{\bf e}_{\boldsymbol{b},\boldsymbol{m}_{\boldsymbol{b}}}\otimes {\bf e}_s,
    \eeq
    where $\{{\bf e}_\alpha\}$ are different sets of Cartesian unit vectors for the readout and state preparation. The columns of this matrix correspond to the relative frequency approximations of the probability vectors encoding the measurement statistics. Here, $N_{s,\boldsymbol{b},\boldsymbol{m}_{\boldsymbol{b}}}$ stands for the number of observed $\boldsymbol{m}_{\boldsymbol{b}}$-outcomes associated to each  measurement basis $\boldsymbol{b}$ and  initial state $\rho_{0,s}$. Hence, $N_{s,\boldsymbol{b}}=\sum_{\boldsymbol{m}_{\boldsymbol{b}}}N_{s, \boldsymbol{b},\boldsymbol{m}_{\boldsymbol{b}}}$ is the number of shots per initialization  and measurement basis, and $N_{\rm shots}=\sum_{s,\boldsymbol{b}}N_{s, \boldsymbol{b}}$ is the total number of shots performed. To estimate the quantum channel, this equation must be inverted using a Moore–Penrose pseudo-inverse, and physical constraints  can be imposed using a maximum-likelihood or least-squares estimation~\cite{PhysRevA.63.020101,PhysRevA.63.054104,PhysRevA.68.012305}.
   
 However, if the noise model under study only contains Pauli channels and the gadget to be characterized only contains Clifford operations, the effective error model will also have the form of a Pauli channel~\eqref{eq:pauliop}, such that the process matrix describing noise will be purely diagonal with $\chi_{ii}=p_i$. If, during simulation, the noisy gadget is prepended by the inverse of the operation implemented by the gadget, such that the operation in absence of noise is the identity, the process matrix obtained from QPT is exactly the process matrix of the error channel. This simplifies the requirements of tomography, as only $3^2=9$ different circuits are required: the qubits need only be initialized in all possible tensor products of the  $+1$ eigenstates of their respective  $X$, $Y$ and $Z$ Pauli operators, and only measured in the same respective basis $\boldsymbol{b}\in\{xx,xy,xz,yx,yy,yz,zx,zy,zz\}$. In this case, the the probability matrix for all the required outcomes contains only $3^2\cdot 2^2=36$ probabilities and is related to the process matrix by the linear map
 \beq
 q_\mu=\sum_{i=0}^{4^m-1}\chi_{ii}\lvert\bra{\mu}P_i\ket{\boldsymbol{b},(+1)^{\otimes m}}\rvert^2
 \label{eq:pauli-tomography-chi}
 \eeq
 where $\ket{\mu}=\ket{\boldsymbol{b},\boldsymbol{m_b}}=\ket{b_1,m_{b_1}}\otimes\cdots\otimes\ket{b_N,m_{b_N}}$ denotes the state after measuring each qubit in the $\boldsymbol{b}$ basis and obtaining outcomes $\boldsymbol{m_b}$, $m=2$ is the number of qubits of the operation being characterized, and we defined $p_0\equiv1-\sum_{i=1}^{4^m-1}p_i$ for convenience. Then, the error channel can be determined by a least-squares inversion with a reduced estimation complexity.

\section{QPT from the detector error model}\label{app:qpt-dem}
We here describe a protocol to extract the measurement outcome probabilities $q_\mu$ from \eqref{eq:pauli-tomography-chi} required to perform QPT of a Clifford circuit under Pauli noise from the DEM of the tomography circuits. The DEM is generated by Stim~\cite{Gidney2021stimfaststabilizer} by tracking the Pauli-frame propagation of each independent error event defined in the noisy circuit and computing a list of measurements (grouped into detectors and logical observables observables) that anti-commute with the error. Thus, the DEM contains a list of $j\in\{1,\dots,N_E\}$ error events, together with its probability $p_j$ and its commutation relation with detectors $\boldsymbol{S}_j=\{+1,-1\}^{\otimes N_\text{detectors}}$ and logical observables $\boldsymbol{L}_j=\{+1,-1\}^{\otimes N_\text{qubits}}$. Detectors are the parity checks for QEC, whose measurement outcomes provide the syndrome $\boldsymbol{S}_j$ to the decoder. In the context of QPT, logical observables correspond to the tomography measurements $(\boldsymbol{b},\boldsymbol{m_b})$. A different DEM is generated for each measurement basis $\boldsymbol{b}$.

Since the DEM lists all independent errors that appear in the circuit, as well as their effect on tomography measurements, it provides enough information to determine the outcome distribution
\begin{equation}
q_{\mu}=\sum_{(i_1,\dots,i_{N_{E,\boldsymbol{b}}})\in\{0,1\}^{N_{E,\boldsymbol{b}}}}\left(\prod_{j=1}^{N_{E,\boldsymbol{b}}}p_{j,\boldsymbol{b}}^{i_j}(1-p_{j,\boldsymbol{b}})^{1-i_j}\right)W_{\mu \boldsymbol{i}}
\end{equation}
where $\boldsymbol{i}=(i_1,...,i_{N_{E,\boldsymbol{b}}})$ runs over all error configurations (i.e. the power set) of the DEM for basis $\boldsymbol{b}$.
The parameter $W_{\mu \boldsymbol{i}}$ determines whether $\boldsymbol{m_b}$ is obtained by the configuration $\boldsymbol{i}$ and is defined as
\begin{equation}
W_{\mu\boldsymbol{i}}=
\begin{cases}
1, & \boldsymbol{m_b}\odot \boldsymbol{C}_{i_1\dots i_{N_{E,\boldsymbol{b}}}}=
\bigodot_{\{j\vert i_j=1\}}\boldsymbol{L}_{j,\boldsymbol{b}}\\
0, & \text{otherwise}.
\end{cases}
\end{equation}
where $\odot$ denotes the element-wise Hadamard product. $\boldsymbol{C}_{i_1\dots i_{N_{E,\boldsymbol{b}}}}$ is a correction vector that flips the raw measurement outcomes $\boldsymbol{m_b}$, and is calculated by the decoder depending on the syndrome associated to error configuration $(i_1,...,i_{N_{E,\boldsymbol{b}}})$. For the bias-filtering CNOT gadget, $\boldsymbol{C}_{i_1\dots i_{2^{d+1}}}=(+1,-1)$ if $b_2\in\{y,z\}$ and more than half of the parity checks are violated, and $(+1,+1)$ otherwise.
The observed syndrome $\boldsymbol{S}_{i_1,...,i_{N_{E,b}}}$ for an error configuration is determined as
\begin{equation}
\boldsymbol{S_i}=\bigodot_{\{j\vert i_j=1\}}\boldsymbol{S}_{j,\boldsymbol{b}}
\end{equation}

For a general DEM, the maximum number of errors is $N_E=2^{N_\text{detectors}+N_\text{qubits}}$, which means that the number of error configurations scales as $2^{2^N}$, and quickly becomes impractical, which makes this protocol interesting only for small circuits. We note that error configurations that include a high number of errors are very unlikely and their contribution to $q_\mu$ is small. Thus, it could be possible to scale this technique by only considering error configurations that contain at most $k$ independent errors. We leave this exploration for future work, as we are interested in a small gadget with $N_\text{qubits}+N_\text{detectors}=4$, where this method works well.

\section{Optimal anisotropy  in the thermodynamic limit}\label{app:optimal-anisotropy}
In this section, we derive an approximate relation between the noise bias $\eta$ of a code capacity channel and the optimal anisotropy of a surface code that maximizes the entanglement fidelity of the logical qubit under this channel. Following~\cite{Benito2025comparativestudyof}, we start by assuming the following dependence between physical and logical error rates
\begin{equation}
\log p_L(d,p)=(a_0+a_1d)\log p+b+b_1d
\label{eq:logical-rate-fit}
\end{equation}
which has been previously shown to  hold consistently for error rates below the respective error  threshold $p_\text{th}$h, which can be approximated as
\begin{equation}
\log p_\text{th}\approx-\frac{b_1}{a_1}.
\label{eq:threshold-fit}
\end{equation}
The entanglement infidelity of the logical qubit is maximized when $p_L(d_X,p/\eta)=p_L(d_Z,p)$, which gives 
\begin{equation}
d_Z=d_X\left(1+\frac{\log\eta}{\log\frac{p_\text{th}}{p}}\right)+\frac{a_1}{a_0}\frac{\log\eta}{\log\frac{p_\text{th}}{p}}
\end{equation}
For sufficiently large system size $N=d_Xd_Z\gg\log\eta$ this equation provides a simple expression for the optimal anisotropy
\begin{equation}
\frac{d_Z}{d_X}\approx1+\frac{\log\eta}{\log\frac{p_\text{th}}{p}}
\label{eq:optimal-anisotropy}
\end{equation}

\begin{figure}
\centering
\includegraphics[width=\linewidth]{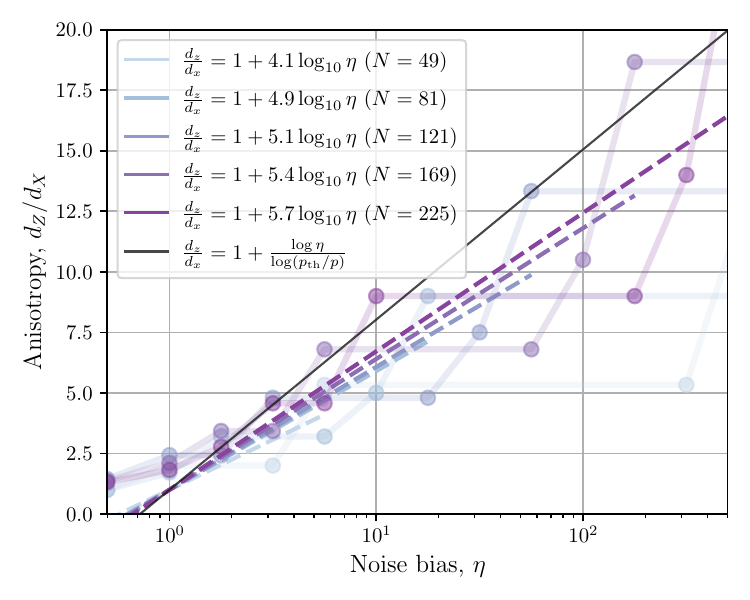}
\caption{{\bf Linear fits for optimal anisotropy}: optimal anisotropy that maximizes the entanglement fidelity of a surface code logical qubit under biased code-capacity noise with $p=0.1$. For every system size $N=d_Xd_Z$, we perform a linear fit to $1+m\log\eta$. Fits for all system sizes have approximately the same slope as predicted by ~\eqref{eq:optimal-anisotropy}, but they differ from the expected scaling (solid black line) due to finite size effects and the threshold approximation of~\eqref{eq:threshold-fit}.}
\label{fig:anisotropy-fits}
\end{figure}

We use that $p_\text{th}=13.9\%$ under our noise model and decoder to plot \eqref{eq:optimal-anisotropy} in Fig.~\ref{fig:anisotropy-fits}. However, the predicted anisotropy does not have a good agreement with data. Because~\eqref{eq:logical-rate-fit} is only valid away from threshold, the approximation in \eqref{eq:threshold-fit} is not fully accurate. Thus, we can keep it as a fitting relation, instead of using its physical interpretation for~\eqref{eq:optimal-anisotropy}. In Fig.~\ref{fig:anisotropy-fits}, we perform a linear fit of the optimal anisotropy of surface codes to  $1+m\log\eta$, where $m$ is the fit parameter. As predicted by \eqref{eq:optimal-anisotropy}, the anisotropy barely depends on the system size when averaging finite-size effects.

\section{XZZX under circuit-level noise}\label{app:xzzx-circ}
\begin{figure}
\centering
\includegraphics[width=\linewidth]{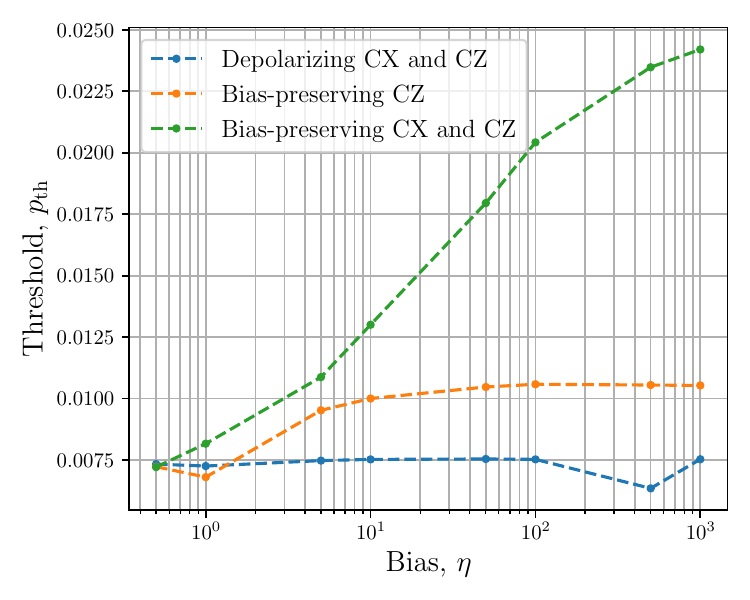}
\caption{{\bf Threshold with native bias-preserving CNOT}: improvement of the XZZX threshold under uniform circuit-level noise ($p=p_\text{2q}=p_\text{id}$, $\eta=\eta_\text{2q}=\eta_\text{id}$) depending on the availability of specific bias-preserving two qubit gates. A bias-preserving CZ can be achieved in experimental platforms, while a bias-preserving CNOT is not available for two-level qubits.}
\label{fig:bias-preserving-cnot}
\end{figure}

In Fig.~\ref{fig:bias-preserving-cnot}, we compare the XZZX code threshold depending on the availability of bias-preserving two-qubit gates with the same performance as their non-bias-preserving counterpart. Under uniform circuit-level noise, and with $\eta>1000$, the threshold is improved by a factor of 3 with respect to unbiased noise if both CNOT and CZ gates are bias preserving, similarly to what happens under biased code-capacity noise. If only CZ are bias-preserving, code threshold is still increased but by just by a $33\%$. Thus, going beyond two-level qubits to implement bias-preserving CNOT gates potentially gives significant performance improvements, provided that such CNOTs can be built with the same reliability as other gates.

\begin{figure}
\centering
\includegraphics[width=\linewidth]{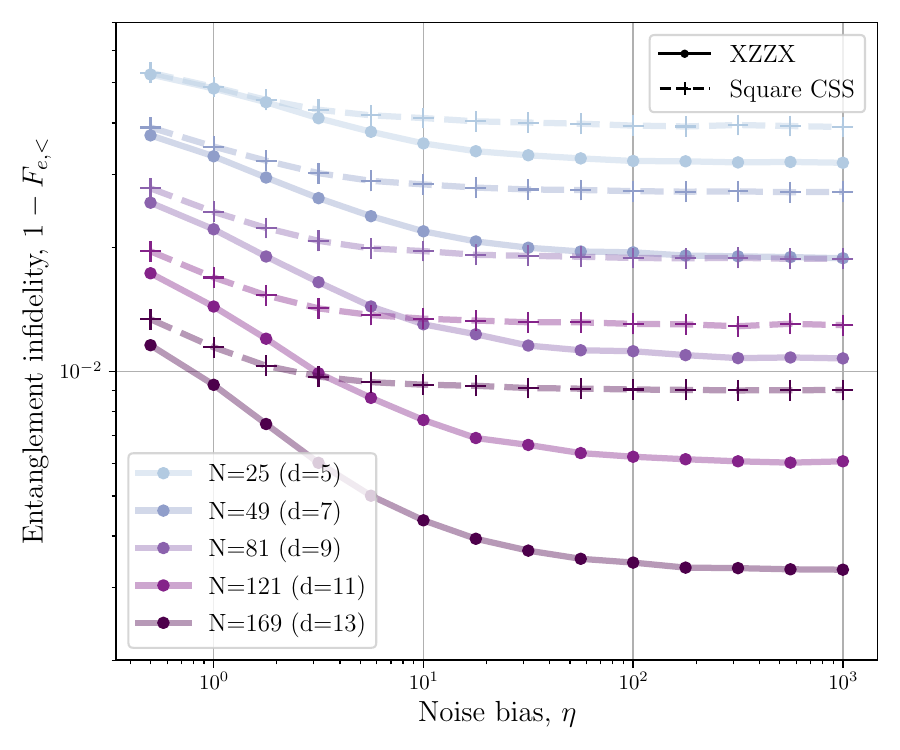}
\caption{{\bf Square CSS vs XZZX code}: comparison of the entanglement infidelity of the square CSS (dashed stars) and XZZX (solid dots) code for a uniformly biased circuit-level noise with $p=p_\text{1q}=p_\text{2q}=p_\text{id}=3\cdot10^{-3}$ and $\eta=\eta_\text{2q}=\eta_\text{id}$. Even though the CNOT gates used in syndrome extraction are not bias-preserving, the remainder bias still provides a significant improvement when using the bias-tailored XZZX code}
\label{fig:square-vs-bias}
\end{figure}

In Fig.~\ref{fig:square-vs-bias}, we compare the square CSS surface code with the XZZX code under uniform biased circuit-level noise. Even if a non-bias-preserving CNOT is used for syndrome extraction, there is a significant performance improvement in the logical error probabilities by using the bias-tailored XZZX code. However, as shown in Fig.~\ref{fig:anisotropic_vs_bias_circuit} of the main text, using anisotropic versions of the CSS surface code provides similar performances to the XZZX code.
\end{document}